\documentclass[twocolumn]{Articles_template_PIC_SRB}  
\usepackage{amsmath}
\usepackage{color}
\usepackage{float}
\usepackage{bm}
\usepackage{hyperref}
\usepackage{soul}
\usepackage{xcolor}
\usepackage{orcidlink}

\begin{document}

\title{Analysis of wave processes using beam-driven Langmuir/$\mathcal{Z}$-mode waveforms generated in Particle-In-Cell simulations}


\author[1]{F. J. Polanco-Rodr\'{i}guez\orcidlink{0009-0005-2951-697X}}
\author[1,2]{C. Krafft\orcidlink{0000-0002-8595-4772}}
\author[1]{P. Savoini\orcidlink{0000-0002-2117-3803}}

\affil[1]{Laboratoire de Physique des Plasmas (LPP), CNRS, Sorbonne Université, 
Observatoire de Paris, Université Paris-Saclay, Ecole polytechnique, Institut Polytechnique de Paris, 91120 Palaiseau, France}
\affil[2]{Institut Universitaire de France (IUF)}

\date{}

\begin{abstract}
    During Type III solar radio bursts, beam-driven upper-hybrid wave turbulence is converted into electromagnetic emissions at the fundamental plasma frequency and its harmonic, through a chain of various linear and nonlinear wave processes. In this work, we mainly investigate the relative roles and interplay of two key mechanisms: the nonlinear decay of Langmuir/$\mathcal{Z}$-mode waves and their linear transformations on random density fluctuations and, in particular, their mode conversion at constant frequency into electromagnetic waves. Using two-dimensional Particle-In-Cell simulations, we employ a  diagnostic approach based on large ensembles of virtual satellites that record local waveforms, enabling detailed temporal and spatial characterization of wave processes in randomly inhomogeneous plasmas. This method allows robust statistical analysis and direct comparison with spacecraft observations. The study focuses on the dependence of wave dynamics on the average level of density fluctuations and the plasma magnetization. Our results quantify the occurrence rate of  decay under varying physical conditions and demonstrate how developed plasma density turbulence can significantly alter the balance between nonlinear wave–wave interactions and linear wave transformations. These findings provide new insights into the mechanisms responsible for electromagnetic emissions during type III radio bursts and strengthen the connection between numerical simulations and in situ solar wind measurements, offering a valuable framework for the interpretation of future space-based waveform observations.
\end{abstract}

\section{Introduction}

Type III solar radio bursts are among the most intense  radio emissions in the solar system (\cite{Dulk1985}, \cite{ReidRatcliffe2014}). Decades of high- and low-frequency waveform analysis have revealed crucial insights about the sequence of processes arising during such bursts  —from the ejection of coronal beams to the radiation of electromagnetic waves  (\cite{Ergun2008}, \cite{MalaspinaErgun2008}, \cite{Malaspina2011}, \cite{GrahamCairns2013b},  \cite{Kellogg2013}, \cite{ThejappaMacDowall2021})—, including more recent  observations  (e.g. \cite{Pulupa2020}, \cite{Pisa2021}, \cite{Soucek2021}, \cite{Larosa2022}, \cite{Formanek2025}, \cite{Pulupa2025}) by the  satellites Solar Orbiter (\cite{Fox2016}) and Parker Solar Probe (\cite{Muller2020}).

During type III bursts, energetic electron beams generate Langmuir and upper-hybrid wave turbulence in the solar wind, which is subsequently converted into electromagnetic radiation at the fundamental plasma frequency  $\omega_p$ and its harmonic $2\omega_p$ via a series of wave-wave, wave-particle and wave-plasma processes. The nonlinear three-wave  electrostatic decay (ESD), where  a Langmuir wave $\mathcal L$ decays into a backscattered wave $\mathcal L'$ and an ion sound wave $\mathcal S'$ through the channel $\mathcal L\rightarrow\mathcal L'+\mathcal S'$ (\cite{Tsytovich1970}, \cite{Melrose1980}), plays a central role in the solar wind. Through the generation of backscattered waves, this process enables the emission of harmonic waves  at  $2\omega_p$ via the  three-wave coalescence mechanism $\mathcal L+\mathcal L'\rightarrow\mathcal H$  (e.g. \cite{Melrose1986},  \cite{Yoon2019}). Additionally, the nonlinear three-wave electromagnetic decay (EMD) can directly contribute  to radio emission at $\omega_p$ through the channel  $\mathcal L\rightarrow\mathcal O+\mathcal S$, where  $\mathcal S$ is an acoustic wave and $\mathcal O$  is the ordinary electromagnetic wave  (\cite{Melrose1980}).

On the other hand, density turbulence is ubiquitous in the solar wind and random density fluctuations $\delta n$  of various wavelengths and amplitudes have been measured (\cite{Celnikier1983,Celnikier1987}, \cite{Kellogg1999b}, \cite{Krupar2018,Krupar2020}). When the typical wavelength of $\delta n$ is much larger than that  of  $\mathcal LZ$  waves, they can interact with them. The strength of these interactions notably depends on the average level of random density fluctuations $\Delta N=\langle (\delta n/n_0)^2\rangle^{1/2}$, the electron beam velocity $v_b$ and the electron plasma thermal velocity $v_T$ ($n_0$ is the  average ambient plasma density). When  $\Delta N\gtrsim 3(v_T/v_b)^2$,  linear  transformations of  Langmuir wave  on density fluctuations  such as reflection, refraction, tunneling, trapping, or conversion are very efficient (\cite{Ryutov1970}, \cite{Krafft2013}). In particular, it was shown that the linear mode conversion (LMC) at constant frequency is the most efficient process of electromagnetic emission at $\omega_p$ in the solar wind (\cite{Krasnoselskikh2019}, \cite{KrafftSavoini2022a}, \cite{Krafft2025}, \cite{KrafftVolokitin2025}). 

Among the various numerical approaches used to study such wave processes, Particle-In-Cell (PIC) simulations of beam-driven Langmuir wave turbulence have shown to be a powerful tool (e.g., \cite{Rhee2009}, \cite{Lee2019}, \cite{ KrafftSavoini2023}, \cite{Polanco2025a}). In this framework, studies have been conducted either regarding nonlinear wave-wave processes in homogeneous or inhomogeneous plasmas (\cite{Kasaba2001}, \cite{Henri2019},  \cite{KrafftSavoini2021,KrafftSavoini2022b}, \cite{Polanco2025a}) or were aimed at understanding  the interactions between wave turbulence and random density fluctuations (\cite{KrafftSavoini2022a}, \cite{Krafft2024}, \cite{Krafft2025}). In a previous work by the authors (\cite{Polanco2025b}), we introduced a new technique using a large number of virtual satellites recording waveforms in a two-dimensional (2D)  PIC simulation plane. Unlike global wave diagnostics, which tend to blend all wave phenomena together, this local approach enables the identification of localized wave processes and their temporal evolution in interaction with other mechanisms. Another advantage is that  such approach mimics actual space-recorded waveforms and facilitates robust statistical analysis of very large sets of simulation data. By directly comparing simulated waveforms with spacecraft observations in the solar wind (e.g. \cite{GrahamCairns2013b}),  we can extend our understanding beyond the physical insights derived  exclusively from space observations  — where many physical quantities cannot be measured simultaneously.  Furthermore, in randomly inhomogeneous plasmas, local characterization of  wave processes is essential because their behavior is highly sensitive to density gradients. 

This work primarily focuses on studying the ESD and LMC mechanisms and assessing their relevance under varying physical conditions, with a particular emphasis on two key parameters : the average level of random density fluctuations, $\Delta N$, and the plasma magnetization ratio, $\omega_c/\omega_p$, where $\omega_c$ is the electron cyclotron frequency. The central objective is to examine the interplay between linear transformations of turbulent  waves on random density fluctuations and  nonlinear wave-wave interaction processes, using detailed waveform analysis.  In this regard, statistical ensembles of waveforms are used to estimate the occurrence rate of the electrostatic decay ---the most efficient nonlinear wave process in the solar wind--- under different physical conditions. This approach allows for the study of its competition with linear mode conversion and the tracking of the temporal evolution of wave turbulence.

Our results provide new perspectives on the wave processes participating in electromagnetic radiation during type III radio bursts, corroborating earlier results obtained through global analysis techniques. By bridging PIC simulations with experimental observations, this work  also establishes a robust foundation for the analysis and interpretation of future space-based waveform data.

\section{Numerical simulations}

Large-scale and  long-term 2D/3V Particle-In-Cell (PIC) simulations are conducted using the SMILEI code (\cite{Derouillat2018}). Simulations cover a computational domain $(x,y)$ of size   $L_x \times L_y = 1448^2\lambda_D^2$ and are performed up to large times $t = 15,000\omega_p^{-1}$, ensuring that the complete development of turbulence is captured. To maintain numerical accuracy over such long time scales and to account for density fluctuations $\delta n$ of a few percent  only of the plasma density, 1800 particles per cell and per species --- plasma ions and electrons as well as electron beam --- are employed. Periodic boundary conditions are used. 

To simulate type III radio bursts' conditions, a weak and energetic electron beam is injected in the simulation plane $(x,y)$ along the background magnetic field $\mathbf{B}_0$  oriented along the $x$-axis. Its drift velocity and relative density are $v_b = 12.7v_T\simeq 0.25c $ --- with $v_T$ denoting the electron thermal velocity --- and $n_b = 5\cdot10^{-4}n_0$ ($n_0$ is the average plasma density), respectively. The mass and temperature ratios between ions and electrons are  $m_e/m_i = 1/1836$ and $T_e/T_i = 10$. When applied initially to the plasma, random density fluctuations $\delta n$ present much larger wavelengths than electrostatic Langmuir and upper-hybrid waves generated by the beam, with average levels up to $\Delta N \leq 0.05$. The cyclotron-to-plasma frequency ratio varies inside the range $0\leq\omega_c/\omega_p \leq 0.14$, corresponding to weakly magnetized plasmas as the solar wind. 

Waveforms are recorded by a large set of virtual satellites moving through the simulation plane with the velocity $v_s=|\mathbf{v}_s|=0.3v_T$ directed along $\mathbf{B}_0$ (\cite{Polanco2025b}). The choice of $v_s$ is relevant to fast solar wind conditions. The six components of electric and magnetic fields, as well as the density of each species, are recorded. The Doppler-shifted frequencies are denoted as $\omega^D=\omega-\textbf{k}\cdot\textbf{v}_s$, where $\omega$ and $\mathbf{k}$ are the  frequency and the wavevectors of waves in the immobile plasma (laboratory) frame. As the ion acoustic frequencies are very small in this frame,  we can write that  $\omega_{\mathcal{S}}^D=\omega_{\mathcal{S}}-\textbf{k}\cdot\textbf{v}_s\simeq -\textbf{k}\cdot\textbf{v}_s$, so that  $\omega_{\mathcal{S}}^D$ can be either negative or positive depending on the direction of propagation of these waves. The ESD three-wave resonance condition is written as  $\omega _{\mathcal{L}%
}=\omega _{\mathcal{L}^{\prime }}+\omega _{\mathcal{S}^{\prime
}} $ in the laboratory frame. In the moving satellite  frame,  it has to be expressed as $\omega _{\mathcal{L}}^{D}=\omega _{\mathcal{L}^{\prime }}^{D}-\omega _{\mathcal{S}^{\prime
}}^{D} $, where the negative sign arises from the Doppler shift associated with the frame transformation.

For additional technical and methodological details, readers are referred to the authors’ previous works  (\cite{KrafftSavoini2021, KrafftSavoini2022a, KrafftSavoini2022b, KrafftSavoini2023, KrafftSavoini2024}, \cite{Krafft2024, Krafft2025}, \cite{Annenkov2025b}, \cite{Polanco2025a, Polanco2025b}).

\section{Homogeneous unmagnetized plasma}\label{section homogeneous}

We begin by analyzing waveforms from simulations conducted in homogeneous and unmagnetized plasmas. The following two sections then explore how plasma density turbulence and magnetization affect the underlying wave phenomena observed in these waveforms. Our focus lies on processes related to electromagnetic wave radiation at $\omega_p$. In this context,  two nonlinear processes are of particular interest in homogeneous and unmagnetized plasmas, i.e. the electromagnetic decay (EMD) which produces  ordinary electromagnetic waves via a direct channel,  and the electrostatic decay, which generates ion acoustic waves capable of triggering the EMD process.

\subsection{Electrostatic decay cascades}
\label{SectionB=0DN=0 exemple cascade}

\begin{figure*}[!htb]
    \centering
    \includegraphics[width=0.95\linewidth]{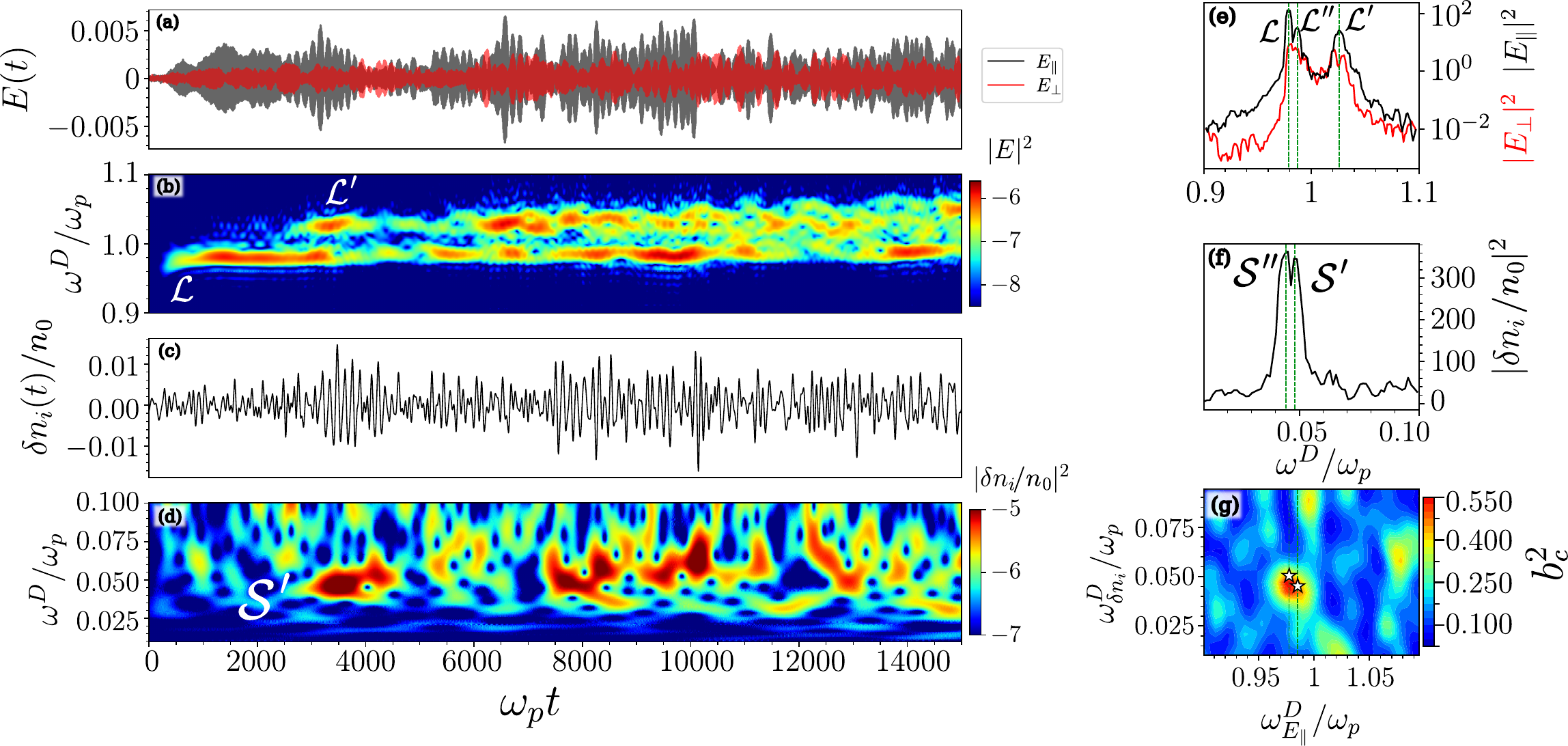}
    \caption{Waveforms in a homogeneous and unmagnetized plasma ($\Delta N=0$, $\omega_{c}=0$). (a) Time variations of the parallel (gray) and perpendicular (red) electric fields $E_{\parallel}(t)$ and $E_{\perp}(t)$. (b) Spectral electric field energy $|E|^{2}$ as a function of the normalized Doppler-shifted frequency $\omega^{D}/\omega_{p}$ and the time $\omega_{p}t.$ (c) Time variations of the ion density perturbation $\delta n_{i}(t)/n_{0}$. (d) Low-frequency spectral energy $|\delta n_{i}/n_{0}|^{2}$  in the map ($\omega^{D}/\omega_{p}$, $\omega_{p}t$). (e)\ High-frequency wave energy spectra $|E_{\parallel}|^{2}$ (black) and $|E_{\perp}|^{2}$ (red), calculated in the time interval $1000\lesssim\omega_{p}t\lesssim6000,$ as a function of $\omega^{D}/\omega_{p}$. (f)\ Low-frequency wave energy spectrum $|\delta n_{i}/n_{0}|^{2}$, in the same time interval and in linear scale, as a function of  $\omega^{D}/\omega_{p}$. (g) Corresponding squared cross-bicoherence $b_{c}^{2}$ calculated for the triplet $(E_{\parallel},\delta n_{i},E_{\parallel})$, in the map ($\omega_{E_{\parallel}}^{D},\omega_{\delta n_{i}}^{D}$); the extrema $b_{c}\simeq0.72$ and $b_{c}\simeq0.68$, represented by stars, appear at $(\omega_{\mathcal{L}}^{D},\omega_{\mathcal{S}^{\prime}}^{D})=(0.978,0.0497)\omega_{p}$ (first cascade) and $(\omega _{\mathcal{L}^{\prime
\prime }}^{D},\omega _{\mathcal{S}^{\prime
\prime }}^{D})=(0.984,0.042)\omega_{p}$ (second cascade), respectively. All variables are in arbitrary units.  }
    \label{fig1}
\end{figure*}

Figures \ref{fig1}a-d present, in a homogeneous unmagnetized plasma, representative waveforms of the parallel and perpendicular electric fields $E_{\parallel }(t)$ and $E_{\perp }(t)$ (a),  the ion density perturbation $\delta n_{i}(t)/n_{0}=(n_i(t)-n_0)/n_0$ (c), as well as the corresponding spectrograms $|E(\omega
^{D},t)|^{2}$ (b) and $|\delta n_{i}(\omega^{D},t)/n_{0}|^{2}$ (d).
The time variation of  $|E(\omega ^{D},t)|^{2}$ shows that beam-driven Langmuir  waves $\mathcal{L}$ are first excited near the Doppler-shifted frequency $\omega _{\mathcal{L}}^{D}\simeq 0.98\omega _{p}$; backscattered $\mathcal{L}^{\prime }$ waves appear later  ($\omega _{p}t\simeq 3000$) at $\omega _{\mathcal{L}^{\prime }}^{D}\simeq 1.03\omega _{p}$, simultaneously with $\mathcal{S}^{\prime }$ waves  excited at $\omega _{\mathcal{S}^{\prime }}^{D}\simeq 0.05\omega _{p}$, showing the first occurrence of the ESD process $\mathcal{L}\rightarrow \mathcal{L}^{\prime }+\mathcal{S}^{\prime }$.
Furthermore, Figure \ref{fig1}e-f display the energy spectra of the Langmuir and ion acoustic waves, calculated within the time interval $1000\lesssim \omega _{p}t\lesssim 6000$, as a function of the  high and low Doppler-shifted frequencies $\omega^D$. One  observes a double-peaked structure corresponding to beam-driven and backscattered Langmuir waves involved in the ESD process, together with a smaller peak representing forward propagating Langmuir $\mathcal{L}^{\prime \prime }$ waves coming from the second decay cascade $\mathcal{L}^{\prime }\rightarrow \mathcal{L}^{\prime \prime }+\mathcal{S}^{\prime \prime}$, as well as two low-frequency  peaks representing the ion acoustic waves $\mathcal{S}^{\prime}$ and $\mathcal{S}^{\prime \prime }$.
Indeed, one can measure that  $\omega _{\mathcal{L}}^{D}\simeq0.978\omega _{p},$ $
\omega _{\mathcal{L}^{\prime }}^{D}\simeq1.026\omega _{p},$ $\omega _{\mathcal{L}^{\prime \prime }}^{D}\simeq0.986\omega _{p},$ $\omega _{\mathcal{S}^{\prime }}^{D}\simeq0.0497\omega _{p}$, and $\omega _{\mathcal{S}^{\prime \prime }}^{D}\simeq0.0420\omega _{p}$.  Then, the three-wave resonance conditions for Doppler-shifted frequencies, i.e. $\omega _{\mathcal{L}%
}^{D}=\omega _{\mathcal{L}^{\prime }}^{D}-\omega _{\mathcal{S}^{\prime
}}^{D} $ ($\omega_{\mathcal{L}}=\omega_{\mathcal{L}^{\prime }}+\omega_{\mathcal{S}^{\prime}}$ in the immobile plasma frame) and $\omega _{\mathcal{L}^{\prime }}^{D}=\omega _{\mathcal{L}^{\prime \prime }}^{D}+\omega _{\mathcal{S}^{\prime \prime }}^{D}$, are  satisfied with reasonable accuracy, as $\left\vert \Delta \omega _{\mathcal{LL}^{\prime }}^{D}\right\vert\ = \left\vert \omega _{\mathcal{L}}^{D}- \omega _{\mathcal{L}^{\prime }}^{D}\right\vert \simeq0.048\omega _{p}$ and  $ \Delta \omega _{{\mathcal{L}}^{\prime}{\mathcal{L}}^{\prime\prime }}^{D} =  \omega _{\mathcal{L}^{\prime}}^{D}- \omega _{\mathcal{L}^{\prime \prime }}^{D} \simeq0.04\omega _{p}$.

Furthermore, the squared cross-bicoherence $b_c^2$, calculated  within the same time interval for the triplet $(E_{\parallel },\delta n_{i},E_{\parallel })$, exhibits the extremum $b_{c}\simeq0.72$ ($b_{c}\simeq
0.68$) for the first (second) cascade (see the stars in Figure \ref{fig1}g and the corresponding caption).  This confirms that the three-wave resonance conditions and phases' coherence between modes  are satisfied, proving the occurrence of two ESD cascades.

We can use the measured frequencies $\omega^D_{\mathcal L}$, $\omega^D_{\mathcal L'}$, $\omega^D_{\mathcal L''}$, $\omega^D_{\mathcal S'}$ and $\omega^D_{\mathcal S''}$     to recover the beam velocity $v_b$, the plasma electron temperature $T_e$ and the ion acoustic velocity $c_s$ (and thus the electron-to-ion temperature ratio) by using the following coupled equations derived in the framework of unmagnetized 1D plasma approximation
\begin{eqnarray}
&\Delta \omega _{\mathcal{LL}^{\prime }}^{D} \simeq \omega _{\mathcal{L}%
}^{D}-\omega _{\mathcal{L}^{\prime }}^{D}\simeq  2\omega _{p}\left( {1}/{v_{b}}-{c_{s}}/{3v_{T}^{2}}%
\right) (c_{s}-v_{s}),  \label{Delta-freq1} \\
&\Delta \omega _{\mathcal{L}^{\prime }\mathcal{L}^{\prime \prime }}^{D}
\simeq \omega _{\mathcal{L}^{\prime }}^{D}-\omega _{\mathcal{L}^{\prime
\prime }}^{D}\simeq 2\omega _{p}\left( {1}/{v_{b}}-{c_{s}}/{v_{T}^{2}%
}\right) (c_{s}+v_{s}),  \label{Delta-freq2} \\
&\Delta \omega _{\mathcal{S}^{\prime }\mathcal{S}^{\prime \prime }}^{D}
\simeq 4c_sv_s/3v_{T}^{2}  ,  \label{Delta-freq3}
\end{eqnarray}
where the ion acoustic frequencies, of the order of a few $10^{-3}\omega
_{p} $ in the plasma (laboratory) frame, have been neglected. 
We have used that $\omega _{\mathcal{L}^{\prime }}=\omega _{\mathcal{L}^{\prime \prime }}+\omega _{\mathcal{S}^{\prime \prime }}$ and $\mathbf{k}_{\mathcal{L}^{\prime }}=%
\mathbf{k}_{\mathcal{L}^{\prime \prime }}+\mathbf{k}_{\mathcal{S}^{\prime
\prime }}$, as well as the  Langmuir and ion acoustic dispersion laws, which leads to $k_{
\mathcal{L}^{\prime \prime }}\simeq k_{b}-2k_{0}$ and $k_{\mathcal{S}^{\prime \prime }}\simeq
-2k_{b}+3k_{0} $ in the 1D  approximation,  where $k_{b}=\omega _{p}/v_{b}$ and $k_{0}\lambda _{D}=2c_{s}/3v_{T}$ (e.g. \cite{Cairns1987}, \cite{KrafftSavoini2024})); $c_{s}$ is the ion acoustic velocity. 
Entering in equations (\ref{Delta-freq1})-(\ref{Delta-freq3})  the above values of $\left\vert \Delta \omega _{\mathcal{LL}^{\prime }}^{D}\right\vert 
$ (or, equivalently, of $\omega _{\mathcal{S}^{\prime }}^{D}$) and $\Delta
\omega _{\mathcal{L}^{\prime }\mathcal{L}^{\prime \prime }}^{D}$, as well as the measured value of  $\Delta \omega _{\mathcal{S}^{\prime }\mathcal{S}^{\prime \prime
}}^{D}\simeq 0.007\omega_p$, we get that $%
v_{b}=13.63v_{T},$ $c_{s}=0.015v_{T},$ and $v_{T}=0.017c$; these quantities are very close to the simulation parameters, i.e. $v_{b}=12.7v_{T},$ $%
c_{s}=0.026v_{T},$ and $v_{T}=0.02c$. Note that if only one decay cascade occurs, only two equations can be used. 

Then, in cases where waves propagate at modest angles relative to the beam direction and in very weakly magnetized solar wind regions near 1 AU, where density turbulence should be weak (i.e. $\Delta N\lesssim 3(v_T/v_b)^2$, \cite{Krafft2013}), high- and low-frequency energy spectra with peaks driven by ESD can be used to diagnose beam and plasma parameters.

\subsection{Dynamics of Langmuir wave turbulence}

\begin{figure}[h]
    \centering
    \includegraphics[width=0.9\linewidth]{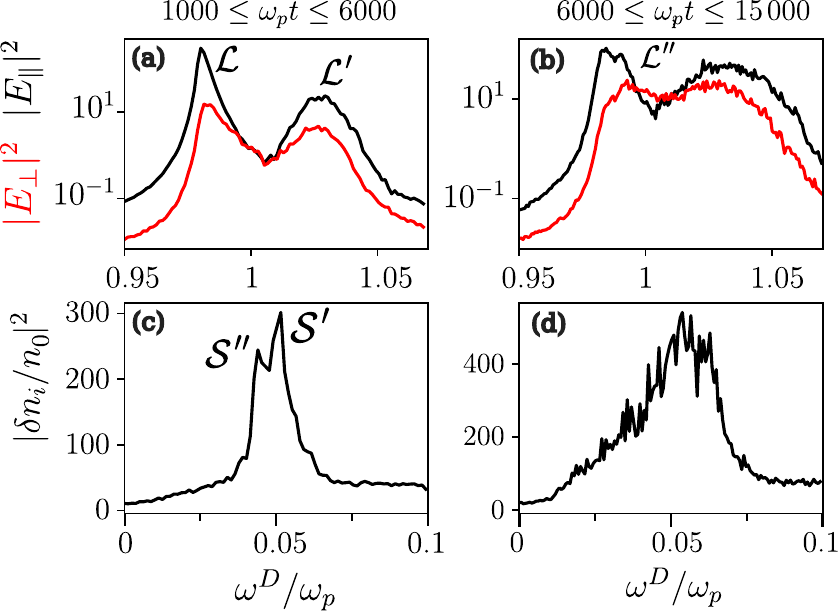}
    \caption{High- and low-frequency energy spectra averaged over $N_s=256$ waveforms, as a function
     of the normalized Doppler-shifted frequency $\omega^{D}/\omega_{p}$. (a-b) Parallel (black) and perpendicular (red) electric field spectra $\left\langle |E_{\parallel}|^{2}\right\rangle $ and $\left\langle |E_{\perp}|^{2}\right\rangle $, in the time intervals $1000\lesssim\omega_{p}t\lesssim6000$ (a) and $6000\lesssim\omega_{p}t\lesssim15,000$ (b).  (c-d) Low-frequency energy spectra $\left\langle |\delta n_{i}/n_{0}|^{2}\right\rangle$,  in the same time intervals as (a) and (b), respectively. (a-b) : logarithmic scales. (c-d) : linear scales. All variables are  in arbitrary units.}
    \label{fig2}
\end{figure}

Figure \ref{fig2} shows the  energy spectra $\left\langle |E_{\parallel
}(\omega^D )|^{2}\right\rangle $,  $\left\langle |E_{\perp }(\omega^D 
)|^{2}\right\rangle $, and  $\left\langle
|\delta n_{i}(\omega^D )/n_{0}|^{2}\right\rangle $, averaged  over a set of $N_w=256$ waveforms, and  calculated in the time intervals $1000\lesssim \omega _{p}t\lesssim 6000$ and $1000\lesssim \omega_{p}t\lesssim 15,000$.  
In the former time range, energy peaks can be identified at frequencies $\omega _{\mathcal{L}}^{D}\simeq0.98\omega _{p},$ $\omega _{\mathcal{L}^{\prime }}^{D}\simeq1.025\omega_{p},$ $\omega _{\mathcal{S}^{\prime}}^{D}\simeq0.052\omega _{p}$, and $\omega _{\mathcal{S}^{\prime \prime }}^{D}\simeq0.044\omega _{p}$ (Figures \ref{fig2}a,c). As expected,  they are very close to those of Figures \ref{fig1}e-f. However, note the larger width of the peak of backscattered  $\mathcal{L}^{\prime}$ waves compared to that of beam-driven ones, due to Doppler-shift effects. One observes  also that $\left\langle|E_{\perp }(\omega^D)|^{2}\right\rangle <\left\langle E_{\parallel }(\omega^D)|^{2}\right\rangle $ for most frequencies, except near $\omega ^{D}\simeq \omega _{p}$ where $\left\langle |E_{\perp }(\omega^D)|^{2}\right\rangle \simeq \left\langle E_{\parallel }(\omega^D
)|^{2}\right\rangle $,  i.e. in the wavevectors' region of electromagnetic waves radiated at $\omega_p$.

For $6000\lesssim \omega _{p}t\lesssim 15,000\ $ (Figures \ref{fig2}b,d), all spectral peaks exhibit significant broadening, attributed to the prolonged occurrence of ESD cascades. Specifically, the second decay cascade becomes evident as a new, smaller peak emerging near $\omega ^{D}\simeq 0.99\omega _{p}$ and corresponding to $\mathcal{L}^{\prime \prime }$ waves. Moreover, the electric energy near $\omega ^{D}\simeq
\omega _{p}$ in Figure \ref{fig2}b is significantly increased, compared to Figure \ref{fig2}a, due to the appearance of Langmuir waves with smaller $k\ll\omega_p/v_b$  involved  in the last stages of ESD and of small-$k$  $\mathcal{O}$-mode waves resulting from the electromagnetic decay  $\mathcal{L}\rightarrow \mathcal{O}+\mathcal{S}$ (\cite{Krafft2024}). On the other hand, the low-frequency spectrum broadens, extending toward both lower and higher frequencies. . Ion acoustic waves with $\omega^D\gtrsim 0.06\omega_p$ ($\omega^D\lesssim0.4\omega_p$) are produced by large-$k$ decaying Langmuir waves generated during the beam deceleration (via EMD and ESD, that are the dominant processes at late times).

\subsection{Resonance conditions and phase coherence between waves}
\begin{figure*}[htb!]
    \centering
    \includegraphics[width=0.85\linewidth]{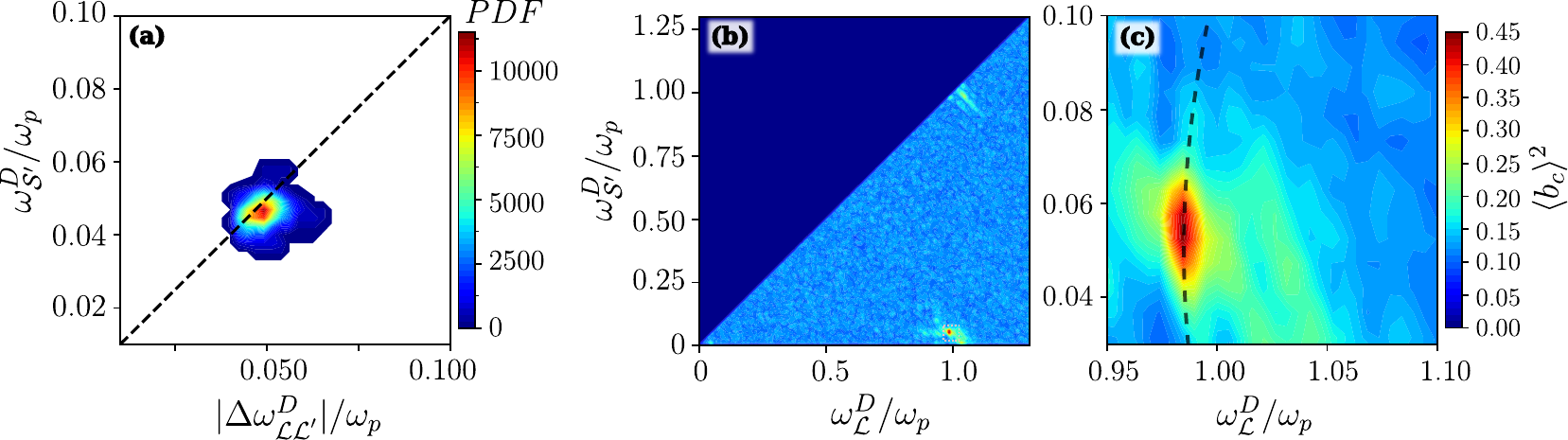}
    \caption{(a) Wave distribution  in the map ($\left\vert\Delta\omega_{\mathcal{LL}^{\prime}}^{D}\right\vert/\omega_p $, $\omega_{\mathcal{S}^{\prime}}^{D}/\omega_p$), obtained  by using $N_s=150 $ selected spectra consistent with ESD occurrence ($3$ peaks identified) out of a set of $256$; the dashed line represents the three-wave resonance condition  $\omega_{\mathcal{L}}^{D}=\omega_{\mathcal{L}^{\prime}}^{D}-\omega_{\mathcal{S}^{\prime}}^{D}$. (b) Squared cross-bicoherence $\langle b_c\rangle^2$ of the triplet ($E_{\parallel},\delta n_{i},E_{\parallel}$), calculated in the time interval $1000\lesssim\omega_{p}t\lesssim6000,$ averaged over $N_{bc}=50$ selected waveforms satisfying at best the resonance condition and consistent with ESD occurrence, in a large  frequency ($\omega_{\mathcal{L}}^{D},\omega_{\mathcal{S}^{\prime}}^{D}$) region. (c) Zoom of (b) in the region $0.95<\omega_{\mathcal{L}}^{D}/\omega_p<1.1$ and $0.03<\omega_{\mathcal{S}^{\prime}}^{D}/\omega_p<0.1$; the dashed line represents the theoretical curve $(\omega _{\mathcal{L}}^{D}(k),\omega _{\mathcal{S}^{\prime }}^{D}(2k-k_0))$ derived using the ESD resonance condition $\omega_{\mathcal{L}}^{D}=\omega_{\mathcal{L}^{\prime}}^{D}-\omega_{\mathcal{S}^{\prime}}^{D}$ and the\ wave dispersion relations. All variables are normalized.}
    \label{fig3}
\end{figure*}

To estimate the fraction of waveforms for which three-wave frequency resonance conditions can be met, $\omega _{\mathcal{S}^{\prime}}^{D}$ is measured as a function of $\left\vert \Delta \omega _{\mathcal{LL}^{\prime}}^{D}\right\vert $ in selected spectra where clear multi-peak structures can be found, as shown in Figure \ref{fig3}a. One can determine that $60\%$ of the analyzed spectra (i.e. $150$ out of $256$) reveal the occurrence of three simultaneous frequency peaks fulfilling the resonance conditions $|\Delta\omega _{\mathcal{LL}^{\prime }}^{D}|\simeq\omega _{\mathcal{L}^{\prime }}^{D}-\omega _{\mathcal{L}}^{D}\simeq\omega _{\mathcal{S}^{\prime}}^{D}$. 
Moreover,  Figures \ref{fig3}b,c present,  in the plane ($\omega_{\mathcal{L}}^{D},\omega_{\mathcal{S}^{\prime}}^{D}$) and for the time interval $1000\lesssim \omega _{p}t\lesssim 6000$, the squared cross-bicoherence $\langle b_{c}\rangle^2$  of the triplet ($E_{\parallel},\delta n_{i},E_{\parallel}$), averaged over the $50$ waveforms presenting high $b_c$-values at frequencies where large spectral energy is recorded --- i.e., compatible with resonant wave-wave interactions.  One observes a maximum $b_c=(\langle b_{c}\rangle^2)^{1/2}\simeq 0.7$ at $\left( \omega _{\mathcal{L}}^{D},\omega _{\mathcal{S}^{\prime}}^{D}\right) \simeq \left( 0.98,0.055\right) \omega _{p},$ which confirms unambiguously that phase coherence between waves is satisfied and thus that ESD indeed occurs for a substantial number of waveforms ($50$ out of $256$, i.e. around $20\%$). Note that no other significant bicoherence level is found outside the frequency region where ESD manifests, as evidenced in Figure \ref{fig3}b. 
The dashed line in Figure \ref{fig3}c represents the theoretical curve derived using the ESD resonance condition $\omega _{\mathcal{L}}^{D}=\omega _{\mathcal{L}^{\prime}}^{D}-\omega _{\mathcal{S}^{\prime }}^{D}$ as well as the Langmuir and ion acoustic dispersion relations, leading to the parametric equation $(\omega _{
\mathcal{L}}^{D}(k),\omega _{\mathcal{S}^{\prime }}^{D}(2k-k_0))$.
A very good coincidence between this theoretical curve and the calculated cross-bicoherence maxima is observed. 

In summary, simulations of an initially homogeneous plasma reveal that wavepackets participating in electrostatic decay are commonly observed. Our analysis indicates that approximately 20\% of these wavepackets fulfill the requirements for both three-wave frequency resonance and phase coherence. Additionally, the occurrence of double electrostatic decay cascades—while not predominant—is documented, suggesting that such multi-step processes do take place under these conditions. 

\section{Randomly inhomogeneous and unmagnetized plasmas}
Density turbulence is ubiquitous in the solar wind. In particular, the transformations of Langmuir and upper-hybrid waves excited by electron beams on random density fluctuations of specific wavelength ranges can generate electromagnetic wave radiation at the plasma frequency. Moreover, when the average level of random density fluctuations meets the condition $\Delta N\gtrsim3(v_T/v_b)^2$  (e.g., \cite{ Ryutov1970}, \cite{Krafft2013}), the presence of such inhomogeneities has a crucial impact on the mechanisms leading to electromagnetic wave emission (e.g., \cite{VolokitinKrafft2018},  \cite{Krasnoselskikh2019}, \cite{KrafftSavoini2022a}). To delve deeper, let us apply the methodology outlined in section \ref{section homogeneous}.

\subsection{Impact  of density fluctuations}

\begin{figure}[H]
    \centering
    \includegraphics[width=\linewidth]{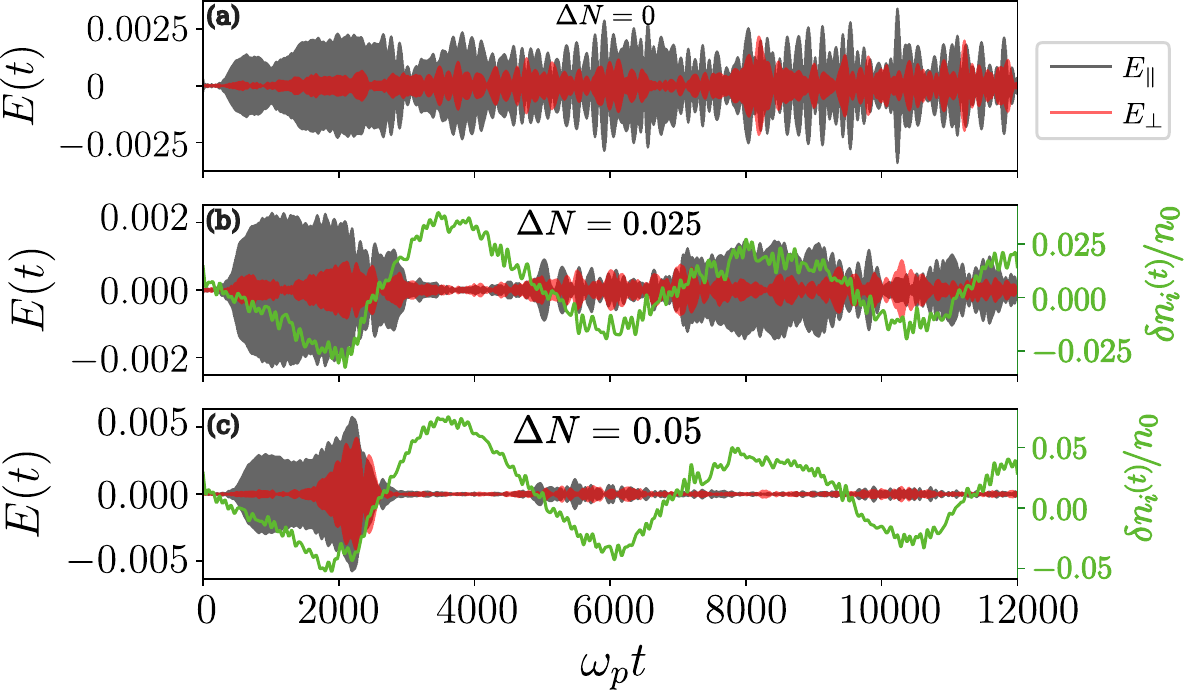}
    \caption{Waveforms of the parallel and perpendicular electric fields $E_{\parallel}(t)$ (gray) and $E_{\perp
}(t)$ (red), to which the time variations of the normalized ion density perturbation $\delta n_{i}(t)/n_{0}$ are superimposed in (b)-(c) (green lines and right axes),  for a plasma with different average levels of density fluctuations : $\Delta N=0$ (a), $\Delta N=0.025$ (b), and $\Delta N=0.05$ (c).  Electric fields are in arbitrary units.}
    \label{fig4}
\end{figure}

Figure \ref{fig4} illustrates typical waveforms of the parallel and perpendicular electric fields, along with the ion density perturbations $\delta n_{i}(t)/n_{0}$,  in plasmas where $\Delta N\lesssim3(v_T/v_b)^2$  ($\Delta N=0$) and $\Delta N\gtrsim3(v_T/v_b)^2$ ($\Delta N=0.025,0.05$). In the homogeneous plasma case ($\Delta N=0$, Figure \ref{fig4}a), the waveform exhibits high-amplitude waves almost continuously, with beat structures lasting $\sim 100\omega_p^{-1}$ that intensify at $\omega_p t\gtrsim3000$, primarily due to the generation of backscattered Langmuir waves via ESD. In the inhomogeneous plasmas (Figures \ref{fig4}b-c), the waveforms show isolated field structures alongside oscillations of significantly reduced amplitudes, tunneling through density humps. Indeed, Langmuir waves undergo reflection, refraction, and trapping in density wells where $\delta n_i<0$ (e.g. \cite{Ergun2008}, \cite{Krafft2014}, \cite{KrafftVolokitin2021}). 
For $\Delta N=0.025$, a wavepacket traverses a density hump ($\delta n_i>0$)   at $7000\lesssim\omega_pt\lesssim10,000$, since its frequency $\omega_{\mathcal{L}}$ exceeds the local plasma frequency $\omega_p(1+\delta n_i/2n_0)$. In contrast, for $\Delta N=0.05$ and in the same time interval, the wavepacket is evanescent when tunneling through a higher density hump where $\omega_{\mathcal L}<\omega_p(1+\delta n_i/2n_0)$. 

While the beats mentioned above are more common in plasmas with $\Delta N\lesssim3(v_T/v_b)^2$,  they can persist even when $\Delta N\gtrsim3(v_T/v_b)^2$, as high amplitude waves are trapped in density wells; in such a case,  forward and backward reflected waves can interact  linearly near reflection points on density gradients, and nonlinear electrostatic decay may also occur locally  (\cite{Krafft2015}, \cite{KrafftSavoini2024}).
For $\Delta N>0$, the most intense waveforms appear at early times (i.e. $500\lesssim\omega_p t\lesssim3000$, Figures \ref{fig4}b-c) and gradually fade away. This attenuation arises as a tail of accelerated beam electrons, formed through wave scattering on density fluctuations, is reabsorbing a significant fraction of  Langmuir wave  energy, leading to its damping (\cite{KrafftSavoini2023}). This effect is most pronounced for large  $\Delta N=0.05$ (see also \cite{Krafft2024}). 

\subsection{ Triggering of electrostatic decay by LMC}
\begin{figure*}[!htb]
    \centering
    \includegraphics[width=\linewidth]{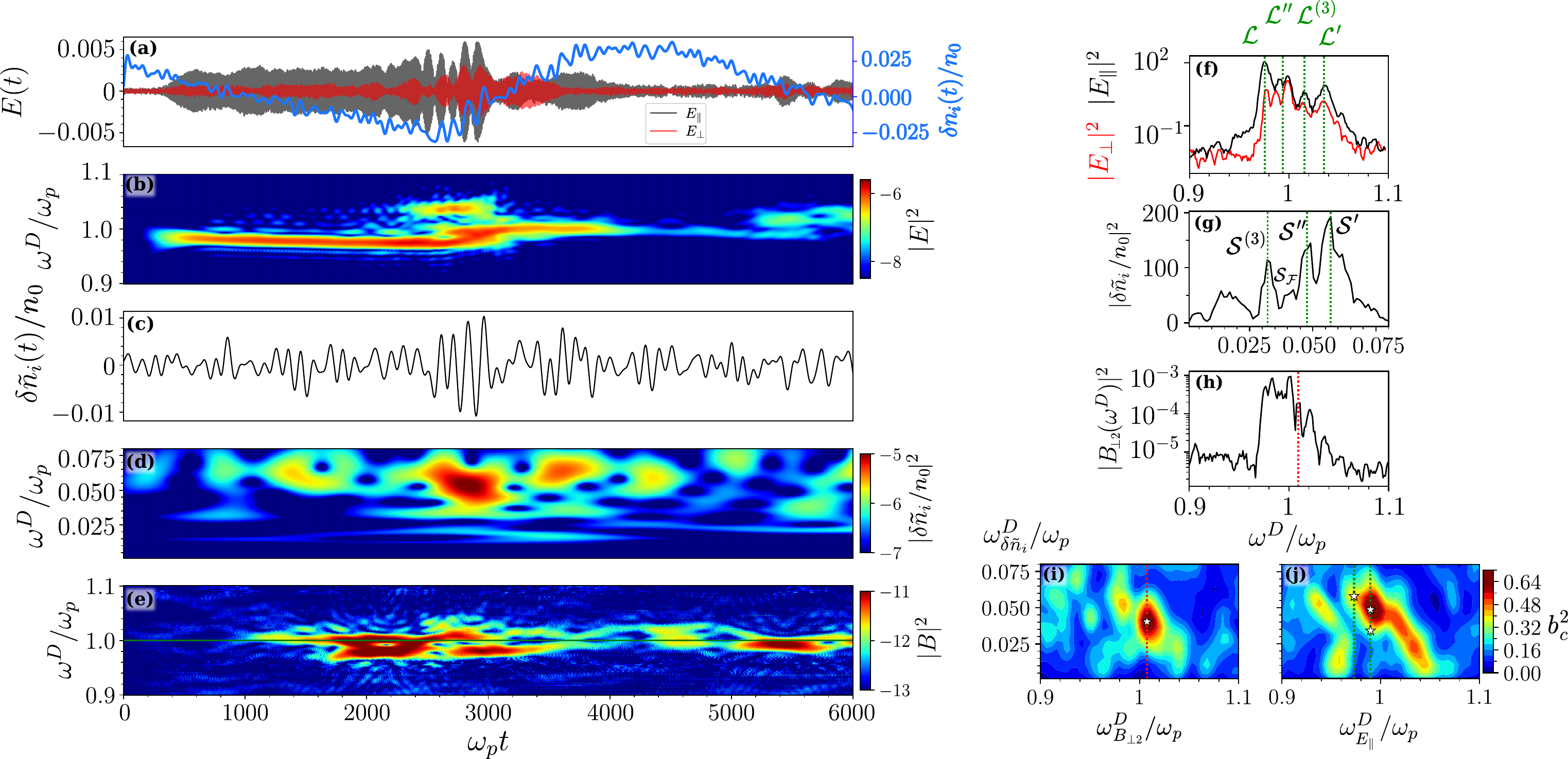}
    \caption{Waveforms in a randomly inhomogeneous  plasma with $\Delta N=0.025$ and $\omega_{c}=0$. (a) Time variations of the parallel (gray) and perpendicular (red) electric fields $E_{\parallel}(t)$ and $E_{\perp}(t)$ (left axis), as well as of the superimposed ion density perturbation $\delta n_i(t)/n_0$ (blue, right axis). (b) Spectral electric field energy $|E|^{2}$ in the map ($\omega_{p}t$, $\omega^{D}/\omega_{p}$). (c) Time variation of the induced ion density perturbation $\delta \tilde{n}_i(t)/n_{0}$ (applied density fluctuations $\delta n(t)$ have been removed from $\delta n_i(t)$ by filtering). (d) Low-frequency spectral energy $|\delta \tilde{n}_{i}/n_{0}|^{2}$  in the map ($\omega_{p}t$, $\omega^{D}/\omega_{p}$). (e) Spectral magnetic field energy $|B_{\perp 2}|^{2}$ in the map ($\omega_{p}t$, $\omega^{D}/\omega_{p}$). (f) High-frequency wave energy spectra $|E_{\parallel}|^{2}$ (black) and $|E_{\perp}|^{2}$ (red) versus $\omega^{D}/\omega_{p}$, calculated in the time interval $\Delta T=[500,6000]\omega_{p}^{-1}$; green labels and vertical lines indicate the excited Langmuir waves and their spectral peaks. (g) Low-frequency wave energy spectrum $|\delta \tilde{n}_{i}/n_{0}|^{2}$ versus $\omega^{D}/\omega_{p}$, calculated within $\Delta T$; green labels indicate the excited ion acoustic waves near their spectral peaks. (h) Magnetic wave energy spectrum $|B_{perp 2}|^2$  versus $\omega^{D}/\omega_{p}$, calculated within $\Delta T$. (i) Squared cross-bicoherence $b_{c}^{2}$ calculated within $\Delta T$ for the triplet $(E_{\parallel},\delta \tilde{n}_{i},E_{\parallel})$, in the map ($\omega_{E_{\parallel}}^{D},\omega_{\delta \tilde{n}_i}^{D}$)$/\omega_p$;   $b_{c}\simeq0.78$ at $(\omega_{\mathcal{L}}^{D},\omega_{\mathcal{S}^{\prime}}^{D})=(0.975,0.058)\omega_{p}$ (first cascade); $b_{c}\simeq0.91$ at $(\omega_{\mathcal{L}^{\prime\prime}}^{D},\omega_{\mathcal{S}^{\prime\prime}}^{D})=(0.99,0.047)\omega_{p}$ (second cascade); $b_{c}\simeq0.67$ at $(\omega_{\mathcal{L}^{\prime\prime}}^{D},\omega_{\mathcal{S}^{(3)}}^{D})=(0.99,0.037)\omega_{p}$ (third cascade); positions in the frequency map are indicated by stars.  (j) Squared cross-bicoherence $b_{c}^{2}$ calculated within $\Delta T$ for the triplet $(B_{\perp 2},\delta \tilde{n}_{i},E_{\parallel})$, in the map ($\omega_{B_{\perp 2}}^{D},\omega_{\delta \tilde{n}_i}^{D}$)$/\omega_p$; $b_{c}\simeq0.78$ at $(\omega_{\mathcal{F}}^{D},\omega_{\mathcal{S}_\mathcal{F}}^{D})=(1.01,0.04)\omega_{p}$, as indicated by  stars.  Parameters are the same as in Figure \ref{fig1}, but with $\Delta N=0.025$. All variables are in arbitrary units.}
    \label{fig5}
\end{figure*}

Figure \ref{fig5} presents, in a plasma with $\Delta N=0.025$, typical waveforms of electromagnetic fields and ion density perturbations, as well as spectral and bicoherence diagnostics. The low- and high-frequency energy spectra $|\delta \tilde{n}_{i}(\omega ^{D})/n_{0}|^{2}$ and $|E(\omega ^{D})|^{2}$ exhibit four and five main peaks, respectively (Figures \ref{fig5}f,g); $\delta \tilde{n}_{i}(t)=\delta n_{i}(t)-\delta n(t)$  is the induced ion density perturbation and $\delta n(t)$ represents the applied density fluctuations that evolve self-consistently.  Green vertical  lines in Figures \ref{fig5}f,g show that wave energy peaks correspond to the frequencies $\omega _{\mathcal{L}}^{D}\simeq0.975\omega _{p},$ $\omega _{\mathcal{L}^{\prime }}^{D}\simeq1.036\omega_{p}$, $\omega _{\mathcal{L}^{\prime\prime }}^{D}\simeq0.99\omega _{p},$ and $\omega _{\mathcal{L}^{(3)}}^{D}\simeq1.017\omega _{p}$ (Langmuir waves), and   $\omega _{\mathcal{S}^{\prime}}^{D}\simeq0.058\omega _{p},$ $\omega _{\mathcal{S}^{\prime\prime }}^{D}\simeq0.047\omega _{p}$, and $\omega _{\mathcal{S}^{(3)}}^{D}\simeq0.033\omega _{p}$ (ion acoustic waves). Then,  the resonance conditions $\omega _{\mathcal{L}^{\prime }}^{D}-\omega _{\mathcal{L}}^{D}=\omega _{\mathcal{S}^{\prime }}^{D}\simeq 0.06\omega _{p},$ $\omega _{\mathcal{L}^{\prime }}^{D}-\omega _{\mathcal{L}^{\prime\prime }}^{D}=\omega _{\mathcal{S}^{\prime\prime }}^{D}\simeq
0.046\omega _{p}$, and $\omega _{\mathcal{L}^{(3)}}^{D}-\omega _{\mathcal{L}
^{\prime\prime }}^{D}=\omega _{\mathcal{S}^{(3)}}^{D}\simeq 0.027\omega _{p}$ for the three cascades of electrostatic decay $\mathcal{L}\rightarrow \mathcal{L}^{\prime }+\mathcal{S}^{\prime }$, $\mathcal{L}%
^{\prime }\rightarrow \mathcal{L}^{\prime \prime }+\mathcal{S}^{\prime
\prime }$ and $\mathcal{L}^{\prime \prime }\rightarrow \mathcal{L}^{(3)}+%
\mathcal{S}^{(3)}$ are fulfilled with reasonable accuracy  (taking into account that the plasma is randomly inhomogeneous). 

Moreover, the squared cross-bicoherence $b_c^2$ calculated  within the time range $1000\leq\omega_pt\leq5500$ (Figure \ref{fig5}i) for the triplet $(E_{\parallel},\delta \tilde{n}_{i},E_{\parallel })$ reaches, around the frequencies listed above, high values for the first ($b_{c}\simeq 0.8$), the second ($b_{c}\simeq 0.9$) and the third ($b_{c}\simeq 0.7$) decay cascades, at positions indicated by stars, confirming the undeniable occurrence of three ESD cascades.
As shown by the waveforms of electric fields and ion density perturbations and their corresponding spectral energies (Figures \ref{fig5}a-d), these cascades occur at the time when beam-driven Langmuir waves with $\omega _{\mathcal{L}}^{D}\simeq0.975\omega _{p}$ reach a reflection point on the gradient of a density fluctuation and interact with the reflected backscattered waves ${\mathcal{L}^{\prime }}$ with frequency  $\omega _{\mathcal{L}^{\prime }}^{D}\simeq1.036\omega_{p}$. Due to their large amplitudes, $\mathcal{L}$ and $\mathcal{L}^{\prime}$ waves can trigger the electrostatic decay process generating ion acoustic waves at $\omega _{\mathcal{S}^{\prime}}^{D}\simeq0.058\omega _{p}$ (Figures \ref{fig5}c-d,i). Meanwhile, ${\mathcal{L}^{\prime\prime}}$ and ${\mathcal{L}^{(3)}}$ waves are excited  at frequencies  $\omega _{\mathcal{L}^{\prime\prime }}^{D}\simeq0.99\omega _{p}$ and $\omega _{\mathcal{L}^{(3)}}^{D}\simeq1.017\omega _{p}$ within the range $2800\lesssim\omega_pt\lesssim3800$ (Figures \ref{fig5}b,f). This shows that wave reflections on random density fluctuations can locally trigger multiple electrostatic decay cascades. 

On the other hand, the appearance of large amplitude magnetic energy at $\omega^D \simeq \omega_p$ within $1800\lesssim\omega_pt\lesssim2500$ is the signature of the linear mode conversion process (LMC) at constant frequency of Langmuir waves $\mathcal L$ into fundamental ordinary electromagnetic waves, hereafter referred to as $\mathcal F$ waves.  This process has been shown to be the most efficient one for generating $\mathcal F$ waves in a plasma with $\Delta N\gtrsim 3(v_T/v_b)^2$ (\cite{KrafftSavoini2022a}, \cite{Krafft2024, Krafft2025}).    
Furthermore, linear mode conversion of Langmuir waves on density fluctuations can trigger the electromagnetic decay (EMD) $\mathcal{L}\rightarrow\mathcal{F}+\mathcal{S_{\mathcal{F}}}$, where $\mathcal{S_{\mathcal{F}}}$  is an ion acoustic wave, due to the simultaneous excitation of $\mathcal{L}$  and $\mathcal F$ waves of significant amplitudes via scattering on $\delta n$ and conversion. This nonlinear process can then occur at much earlier times than in a homogeneous plasma.  It is likely responsible for the  magnetic energy $|B_{\perp 2}(\omega ^{D},t)|^{2}$ observed  at $\omega^D\gtrsim\omega_p$ within $2500\leq\omega_pt\leq 3500$ (Figure \ref{fig5}e).  Remind that, in an unmagnetized plasma, the  field component $B_{\perp 2}=B_z$ (perpendicular to the two-dimensional simulation plane $(x,y)$), dominates the magnetic energy, which is primarily carried by ordinary wave modes. The cross-bicoherence  calculated in Figure \ref{fig5}j by using the triplet $(B_{\perp 2},\delta \tilde{n}_i,E_{\parallel})$  shows that the maximum $b_c\simeq0.8$  is reached at $\omega^D_{\mathcal S_{\mathcal F}}\simeq0.035\omega_p$ and $\omega^D_{\mathcal F}=1.01\omega_p$. 
This confirms that  electromagnetic waves emitted at $\omega^D\gtrsim\omega_p$ originate from the EMD of  ${\mathcal{L}^{\prime }}$ waves with frequencies  $\omega^D_{{\mathcal{L}^{\prime }}}\simeq1.045\omega_p$, via the channel $\mathcal{L'}\rightarrow \mathcal{F}+\mathcal{S_{\mathcal{F}}}$ (see also  Figure \ref{fig5}f). This result is consistent with the authors' earlier findings  (\cite{Krafft2024}), which were derived using a global approach (as opposed to the local method employed here). Furthermore, the linear transformation of $\mathcal L$ waves on $\delta n$ might also stimulate the coalescence $\mathcal L+\mathcal L'\longrightarrow\mathcal H$ like it does with ESD and EMD. Indeed, \cite{KrafftSavoini2021} observed a faster  $\mathcal H$-wave emission in plasmas where $\Delta N\gtrsim3(v_T/v_b)^2$.

The magnetic signature observed  at $\omega^D\lesssim\omega_p$, simultaneously  with EMD,  corresponds likely to  $\mathcal F$ waves generated  via linear mode conversion  near the reflection point on the density gradient at $\omega_pt\simeq 2800$, which can stimulate EMD. Similarly, as previously exposed, the linear transformations of $\mathcal L$ waves on density fluctuations can trigger ESD as well, by generating backscattered $\mathcal L^{\prime}$ waves of significant amplitudes. Then, via the growth of $\mathcal S^{\prime}$ waves with wavevectors close to those of ion acoustic waves  $\mathcal S_{\mathcal{F}}$ involved in EMD,  this process, in turn, can also be stimulated (\cite{Krafft2024}). Such kind of electromagnetic wave generation at density gradients following the occurrence of LMC is frequently observed in the waveforms recorded in our simulations with randomly inhomogeneous plasmas, which confirms that LMC can trigger nonlinear phenomena at early times, as suggested previously by the authors (\cite{Krafft2024}). 

\subsection{Occurrence of electrostatic decay in plasmas with density turbulence}

\begin{figure}[htb!]
    \centering
    \includegraphics[width=\linewidth]{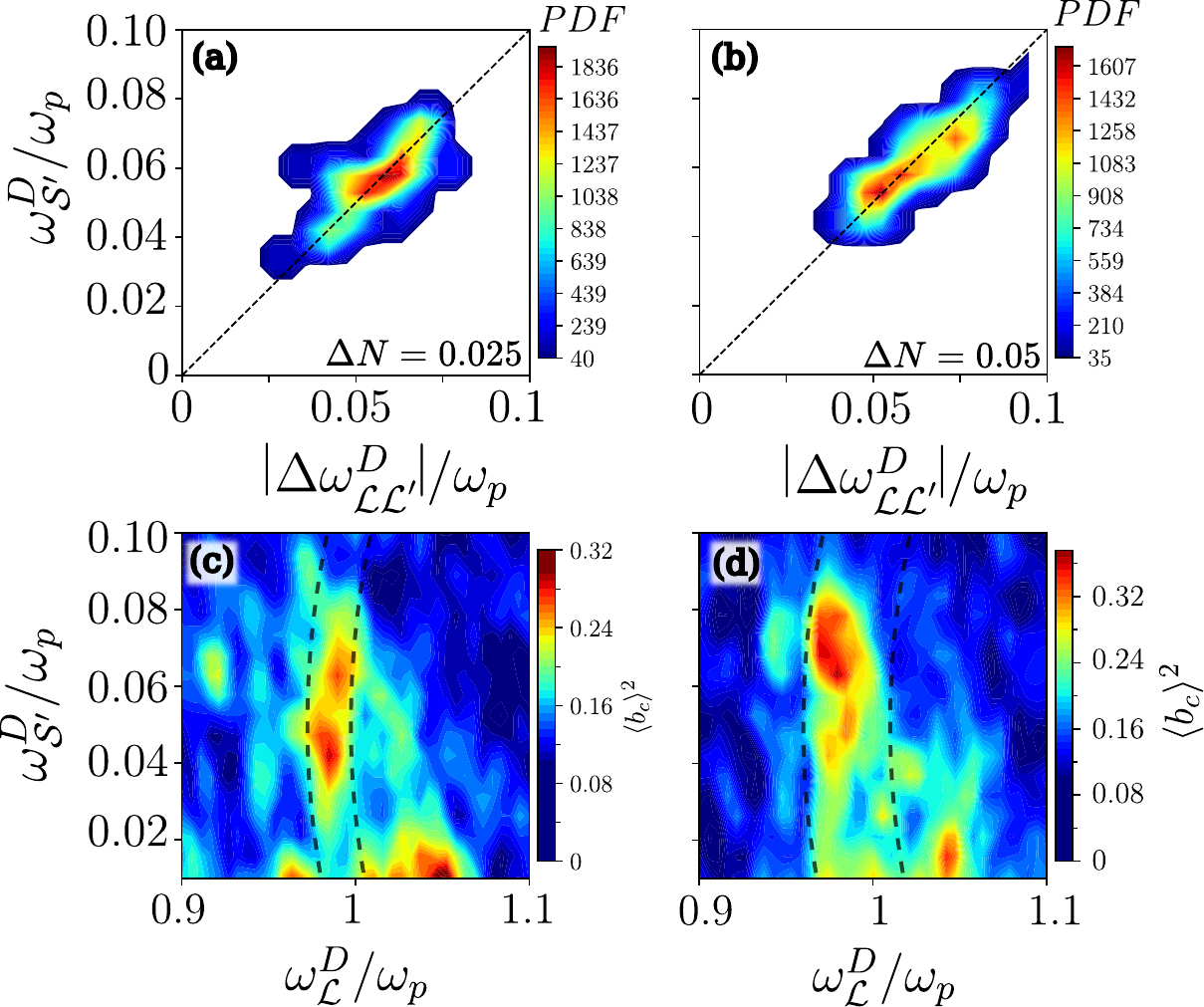}
    \caption{(a-b) Wave distributions in the plane ($\left\vert\Delta\omega_{\mathcal{LL}^{\prime}}^{D}\right\vert/\omega_p $, $\omega_{\mathcal{S}^{\prime}}^{D}/\omega_p $), obtained using  $N_s=60 $ ($N_s=59 $) selected spectra featuring  a multi-peak structure, out of a set of $256 $, for $\Delta N=0.025$  and $\Delta N=0.05$, respectively; the  dashed lines represent the three-wave resonance condition $\omega _{\mathcal{L}}^{D}=\omega_{\mathcal{L}^{\prime }}^{D}-\omega _{\mathcal{S}^{\prime }}^{D}$. (c-d) Squared average cross-bicoherence $\langle b_c\rangle^2$ of the triplet ($E_{\parallel},\delta \tilde{n}_{i},E_{\parallel}$), calculated in the time interval $1000\lesssim\omega_{p}t\lesssim6000,$ in the Doppler-shifted frequency map $(\omega_{\mathcal{L}}^{D}/\omega_p , \omega_{\mathcal{S}^{\prime}}^{D}/\omega_p )$, for $\Delta N=0.025$ ($\Delta N=0.05$); averaging is done over $N_{bc}=20$ ($N_{bc}=13$) selected spectra consistent with ESD occurrence, out of 256; the dashed lines represent the parametric curves $(\omega_{\mathcal L}^D(k)+\alpha\Delta N\omega_p/2,\omega_{\mathcal S'}^D(2k-k_0))$ for $\ \alpha =-1,1$,  derived using the resonance condition $\omega_{\mathcal{L}}^{D}=\omega_{\mathcal{L}^{\prime}}^{D}-\omega_{\mathcal{S}^{\prime}}^{D}$ and the\ wave dispersion relations. Simulation parameters are the same as in Figure  \ref{fig1}, but with  $\Delta N>0$. All variables are normalized.}
    \label{fig6}
\end{figure}

Figures \ref{fig6}a,b display wave distributions in the map ($\left\vert \Delta \omega _{\mathcal{LL}^{\prime}}^{D}\right\vert/\omega_p $, $\omega _{\mathcal{S}^{\prime }}^{D}/\omega_p$), obtained by using selected energy spectra featuring a multi-peak structure, in  plasmas with $\Delta N=0.025$ (a) and $\Delta N=0.05$ (b).
The  distributions are scattered along the ESD frequency resonance condition $|\Delta \omega_{\mathcal{LL}'}|=\omega _{\mathcal{L}^{\prime }}^{D}-\omega _{\mathcal{L}}^{D}=\omega_{\mathcal S'}$ (represented by dashed lines), which is hardly satisfied because of wave scattering on  density fluctuations, but not prohibited locally (\cite{KrafftSavoini2024}). More precisely, around  $20\%$ of the spectral peaks satisfy the resonance condition (see the caption).  Compared to the homogeneous plasma case, the distributions are noticeably shifted toward larger ion acoustic frequencies, likely due to the presence of scattered  Langmuir  waves  with  larger $k$. Additionally, the frequency band $\left\vert \Delta \omega _{\mathcal{LL}^{\prime}}^{D}\right\vert $ widens with  increasing $\Delta N$, consistent with the frequency broadening induced by the density fluctuations $\Delta N\omega_p$.

Figures \ref{fig6}c,d present the squared cross-bicoherence $\langle b_c\rangle^2$ averaged over the  $5-10\%$ of waveforms (see the caption) consistent with electrostatic decay. 
The bicoherence maxima are bounded by the parametric curves $(\omega_{\mathcal L}^D(k)+\alpha\Delta N\omega_p/2,\omega_{\mathcal S'}^D(2k-k_0))$ with $\ \alpha =-1,1$ (dashed lines).
The extrema $\langle b_c\rangle\simeq0.54$  and   $\langle b_c\rangle\simeq0.6$ are located at $(\omega_{\mathcal L}^D,\omega_{\mathcal S'}^D)\simeq(0.98,0.055)\omega_p$ for $\Delta N=0.025$ (c) and $(\omega_{\mathcal L}^D,\omega_{\mathcal S'}^D)\simeq(0.97,0.065)\omega_p$ for $\Delta N=0.05$ (d), respectively; lowest frequencies $\omega_{\mathcal L}^D\simeq0.97\omega_p$  correspond to Langmuir waves trapped in density depletions.
Once again, one observes that Doppler-shifted frequencies of ion acoustic and Langmuir waves are widespread, due to significant wave scattering on density fluctuations.
The highest levels of bicoherence correspond roughly to similar frequencies $(\omega_{\mathcal L}^D,\omega_{\mathcal S'}^D)$ for both $\Delta N$, but not exactly.  Indeed,  ESD occurs only in localized plasma  regions where density turbulence is sufficiently weak not to destroy coherence between waves, what depends on $\Delta N$. 
Finally, note that $\langle b_c\rangle\simeq0.5$ at $(\omega_{\mathcal F}^D,\omega_{\mathcal S_{\mathcal{F}}}^D)\simeq(1.02, 0.016)\omega_p$  (Figures \ref{fig6}c,d), indicating the occurrence of electromagnetic decay, stimulated by LMC. 

\subsection{Spectral characteristics of Langmuir wave turbulence}
\begin{figure*}[!htb]
    \centering
    \includegraphics[width=0.7\linewidth]{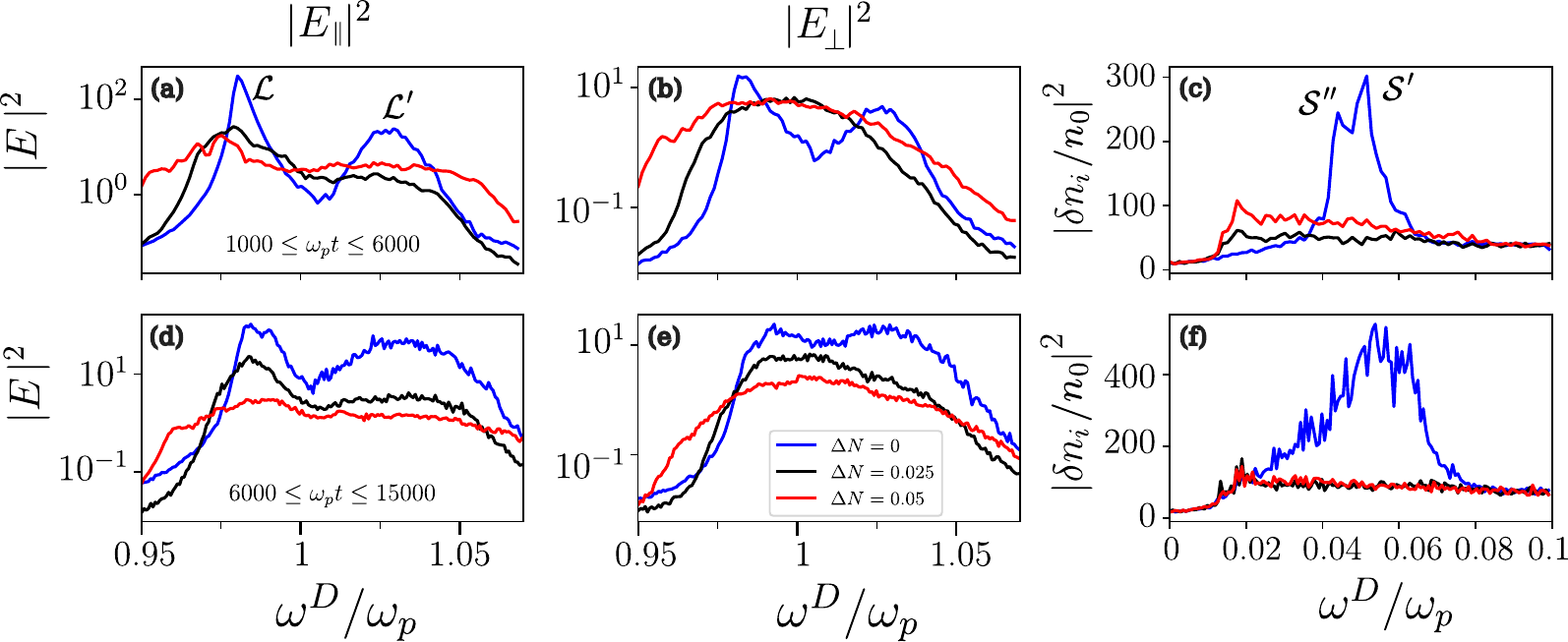}
    \caption{Energy spectra averaged over $256$ waveforms in the time range $\Delta T$, as a function of the Doppler-shifted  frequencies $\omega^{D}/\omega_{p}$, for $\Delta N=0$ $\ $(blue), $\Delta N=0.025$ (black), and $\Delta N=0.05$ (red). (a-c) Parallel $\left\langle| E_{\parallel}|^{2}\right\rangle $ and perpendicular $\left\langle |E_{\perp}|^{2}\right\rangle $ electric field spectra,  as well as ion acoustic energy spectrum $\left\langle |\delta \tilde{n}_{i}/n_{0}|^{2}\right\rangle $, for $\Delta T= [1000,6000]\omega_p^{-1}$. (d-f)  $\left\langle| E_{\parallel}|^{2}\right\rangle $,  $\left\langle |E_{\perp}|^{2}\right\rangle $, and $\left\langle |\delta \tilde{n}_{i}/n_{0}|^{2}\right\rangle $,  for $\Delta T= [6000, 15,000]\omega_p^{-1}$.  All variables are  in arbitrary units.}
    \label{fig7}
\end{figure*}

Figures \ref{fig7}a-f  show the high- and low-frequency averaged  spectra of $\left\langle |E_{\parallel }|^{2}\right\rangle $, $\left\langle |E_{\perp }|^{2}\right\rangle $, and $\langle|\delta \tilde{n}_{i}/n_0|^{2}\rangle$ versus  $\omega^{D}/\omega_{p}$,  for $\Delta N=0,0.025$ and $0.05$, in the  time ranges $1000\lesssim \omega _{p}t\lesssim6000$ (a-c) and $6000\lesssim \omega _{p}t\lesssim15,000$ (d-f). 
At early times (see  Figures \ref{fig7}a-c at $ \omega _{p}t\lesssim 6000)$, when $\Delta N$ increases  from 0 to $0.05$, the energy $\left\langle |E_{\parallel}|^{2}\right\rangle $  of beam-driven $\mathcal{L}$  (backscattered  $\mathcal{L}^{\prime }$) waves decreases (tends to a quasi-flat scattered distribution), as a result of wave scattering on density inhomogeneities and damping by beam reabsorption. Random density fluctuations are responsible for the  broadening of  $\left\langle |E_{\parallel}|^{2}\right\rangle $ and $\left\langle |E_{\perp }|^{2}\right\rangle $, which increases with $\Delta N$.
For $\Delta N=0$, the significant increase with time  of   $\left\langle |E_{\parallel }|^{2}\right\rangle $ and $\left\langle |E_{\perp }|^{2}\right\rangle $ within the frequency range  $0.99\lesssim\omega^D/\omega_p\lesssim1.01$ results from Langmuir wave energy transport to small $k$ through nonlinear processes as ESD and EMD.

 For $\Delta N>0$, $\left\langle |\delta
\tilde{n}_{i}/n_0|^{2}\right\rangle $ is strongly flattened in both early and late time intervals, showing the predominance of  Langmuir  wave  transformations on density fluctuations and a significant reduction of  three-wave interaction processes as ESD and EMD. The extension of $\left\langle |\delta \tilde{n}_{i}/n_0|^{2}\right\rangle $ down to  $\omega^{D}\lesssim 0.02\omega _{p}$ is explained by (i) the occurrence of EMD  stimulated  at early stages  by LMC,  producing ion acoustic waves of lower frequencies than those generated by ESD, and (ii) the fact that $\mathcal L$ waves scattered on $\delta n$ produce,  through ESD and EMD, oblique ion acoustic waves with  smaller frequencies $\omega^D(k)\simeq |kv_s\cos\theta|$.
Moreover, due to Langmuir wave scattering and its impact on three-wave resonance conditions,  the distribution $\left\langle |\delta \tilde{n}_{i}/n_0|^{2}\right\rangle $ extends over a larger frequency  range than for $\Delta N=0$ (compare with  Figure \ref{fig2}). Note that while high levels of cross-bicoherence are reached for  $0.04\lesssim\omega^D/\omega_p\lesssim0.07$ (Figures \ref{fig6}c,d),  $\langle |\delta \tilde{n}_i/n_0|^2\rangle$ peaks at $\omega^D\simeq0.02\omega_p$. This can be attributed to two main factors. First, the Doppler-shifted frequencies of ion acoustic waves involved in ESD are broadly distributed (Figures \ref{fig6}a-d); second, most analyzed waveforms lack a very  distinct ESD signature. As a result, when averaged, these waves contribute only limited statistical significance to the ESD process.

In addition,  Figures \ref{fig7}d-f show the corresponding distributions within a later time interval  $6000\lesssim\omega _{p}t\lesssim 15,000$. One observes that all spectra are significantly  broadened, due to ESD and EMD processes (linear  transformations of Langmuir waves on $\delta n$ for $\Delta N=0$ ($\Delta N>0$). For $\Delta N>0$ ($\Delta N=0$), both electric energy spectra show that Langmuir wave turbulence is strongly (weakly) damped, due to energy reabsorption by the beam (\cite{KrafftSavoini2023}). When $\Delta N>0$, the  distributions' shapes differ significantly from those observed at earlier times (Figure \ref{fig7}c). Indeed, at $\omega_pt\gtrsim6000$, Langmuir waves are already substantially damped, and intense wavepackets become  rare, as well as wave-wave interactions.

In summary, our analysis demonstrates that electrostatic decay (ESD) persists in localized regions of randomly inhomogeneous plasmas. Over 20\% of the observed spectra exhibit three-wave resonant interactions between high- and low-frequency spectral peaks, while 7\% retain phase coherence within wave triplets. The Langmuir and ion acoustic wave spectra, however, appear more scattered and diffuse, with diminished peak intensities at high frequencies. This shows that linear transformations of Langmuir waves on density fluctuations dominate, particularly through the LMC process, which efficiently converts Langmuir waves into electromagnetic waves at constant frequency. Despite reduced phase coherence, decay processes involving multiple cascades endure in inhomogeneous plasmas. Notably, we provide clear evidence of LMC stimulating both EMD and ESD, as well as ESD driving EMD.

By adopting a local approach, this work corroborates and extends findings previously identified or suggested through global analysis  (\cite{Krafft2024}), offering however complementary insights into the interplay between  nonlinear interactions between waves and  linear transformations of waves in randomly inhomogeneous plasmas. 

 \section{Weakly magnetized homogeneous plasmas}

In magnetized plasmas,  the beam-plasma instability generates  upper-hybrid wave turbulence. The excited quasi-electrostatic waves are also called magnetized Langmuir or Langmuir/$\mathcal Z$-mode waves, and are below referred to as $\mathcal{LZ}$ waves. In weakly magnetized homogeneous plasmas, the dynamics of turbulent $\mathcal{LZ}$ waves is mainly dominated by nonlinear wave-wave processes, on which the  impact of plasma magnetization is studied below.

\subsection{Impact of plasma magnetization on wave processes}\label{SectionB=007DN=0 exemple cascade}

\begin{figure*}[!htb]
    \centering
    \includegraphics[width=\linewidth]{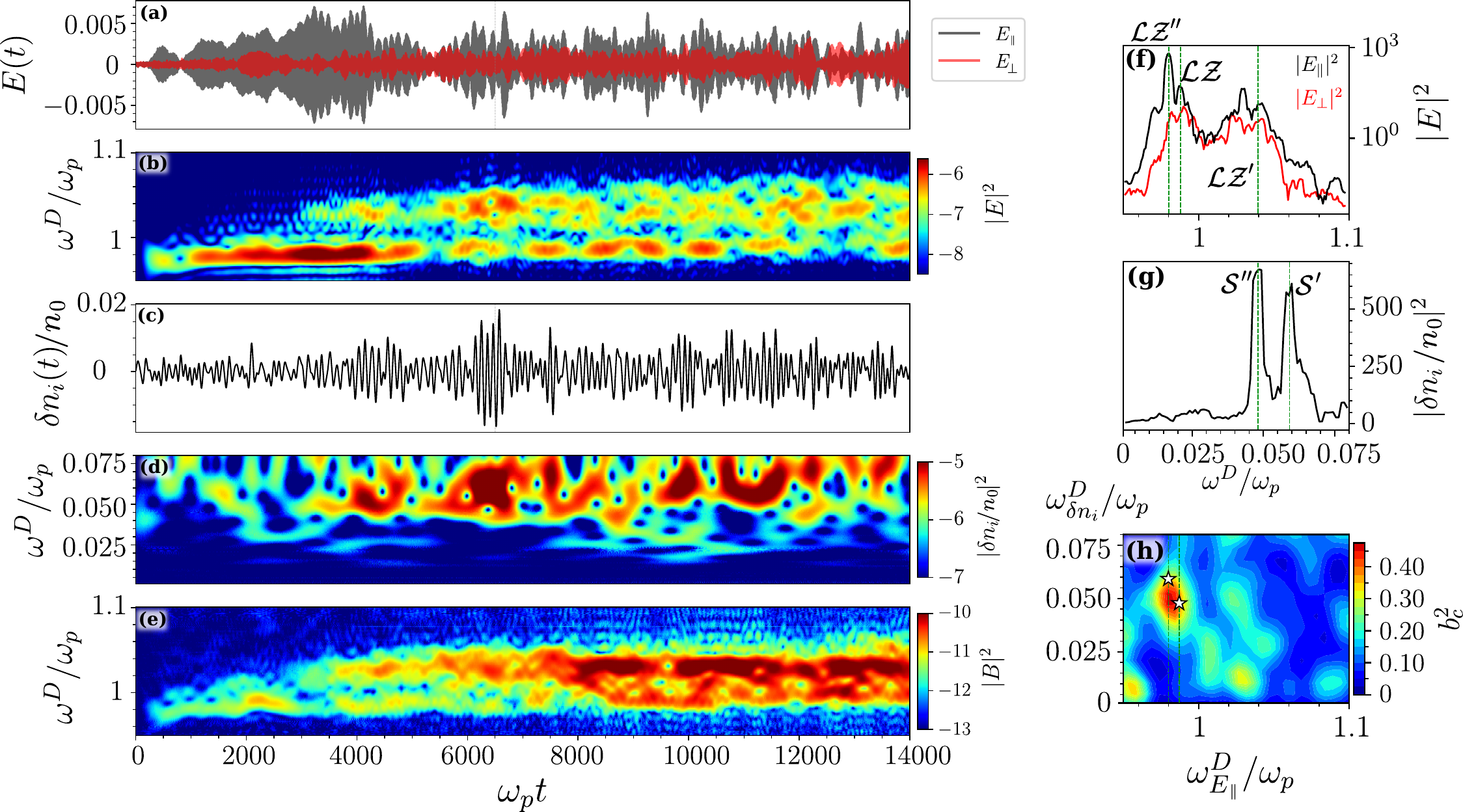}
    \caption{Waveforms in a homogeneous and weakly magnetized plasma with $\omega_{c}/\omega_{p}=0.07$. (a) Time variations of the parallel (gray) and perpendicular (red) electric fields $E_{\parallel}(t)$ and $E_{\perp}(t)$. (b) Spectral electric field energy $|E|^{2}$ in the map ($\omega_{p}t$, $\omega^{D}/\omega_{p}$). (c) Time variation of the  ion density perturbations $\delta n_i(t)/n_{0}$. (d) Low-frequency spectral energy $|\delta n_{i}/n_{0}|^{2}$  in the map ($\omega_{p}t$, $\omega^{D}/\omega_{p}$). (e) Spectral magnetic field energy $|B|^{2}$ in the map ($\omega_{p}t$, $\omega^{D}/\omega_{p}$).
    (f) High-frequency wave energy spectra $|E_{\parallel}|^{2}$ (black) and $|E_{\perp}|^{2}$ (red), versus $\omega^{D}/\omega_p$, calculated in the time interval $\Delta T=[0,6500]\omega_{p}^{-1}$; green vertical lines indicate the spectral peaks of excited $\mathcal{LZ}$ waves. (g) Low-frequency wave energy spectrum $|\delta n_{i}/n_{0}|^{2}$ versus $\omega^{D}/\omega_p$, calculated within $\Delta T$; green vertical lines indicate the spectral peaks of excited ion acoustic waves. (h) Squared cross-bicoherence $b_{c}^{2}$ calculated within $\Delta T$ for the triplet $(E_{\parallel},\delta n_{i},E_{\parallel})$, in the map ($\omega_{E_{\parallel}}^{D}/\omega_p, \omega_{\delta n_i}^{D}/\omega_p$); $b_{c}\simeq0.6$ at $(\omega_{\mathcal{LZ}}^{D},\omega_{\mathcal{S}^{\prime}}^{D})=(0.979,0.059)\omega_{p}$ (first cascade); $b_{c}\simeq0.63$ at $(\omega_{\mathcal{LZ}^{\prime\prime}}^{D},\omega_{\mathcal{S}^{\prime\prime}}^{D})=(0.987,0.048)\omega_{p}$ (second cascade); positions are indicated  in the map by stars. All variables are in arbitrary units.}
    \label{fig8}
\end{figure*}

In a weakly magnetized plasma with $\omega_{p}\gg \omega _{c},$  $\mathcal{LZ}$ waves follow at small $k$ the dispersion of the slow-extraordinary mode, the so-called electromagnetic  $\mathcal{Z}$-mode with the cutoff frequency $\omega\simeq\omega_p-\omega_c/2$. 
The transition between the Langmuir-like and the $\mathcal{Z}$-mode-like wave dispersions occurs along the parallel direction around the wavenumber $k_{\ast }\lambda _{D}=(v_{T}/c)(1+\omega _{p}/\omega_{c})^{-1/2}$. Other electromagnetic modes exist at $\omega\simeq\omega_p$, i.e. the ordinary $\mathcal O$-mode and the fast-extraordinary $\mathcal X$-mode, with the cutoff frequencies $\omega=\omega_p$  and  $\omega\simeq\omega_p+\omega_c/2$, respectively.

Figures \ref{fig8}a-e show representative  waveforms of the parallel and perpendicular electric fields, together with the ion density perturbation and the corresponding spectrograms of  $|E|^{2}$, $|B|^{2}$ and $|\delta n_{i}/n_{0}|^{2}$. 
At early times $\omega_pt\lesssim 4000$,  the parallel field component $E_\parallel(t)$  dominates over the perpendicular one,  $E_\perp(t)$.  As time progresses, the amplitude of $E_\perp(t)$ grows, and wave beatings emerge around $\omega_pt\gtrsim3500$, jointly with backscattered $\mathcal{LZ^\prime}$ and ion acoustic  $\mathcal{S^\prime}$ waves (Figures \ref{fig8}b-d). Jointly, magnetic energy appears at $\omega^D\gtrsim\omega_p$, indicating the generation of $\mathcal{LZ}$ ($\mathcal{O}$-mode)  waves produced through electrostatic (electromagnetic) decay. Later, at $\omega_pt\gtrsim8500$,  $\mathcal Z$-mode waves are generated at frequencies $\omega^D\lesssim\omega_p$ during the ultimate stage of ESD (\cite{Polanco2025a}).

The energy spectra of $\mathcal{LZ}$ and ion acoustic waves calculated in the range $\omega _{p}t\lesssim 6500$ are shown in Figures \ref{fig8}f-g as a function of $\omega^{D}/\omega_{p}$. One observes the signatures of the two decay cascades $\mathcal{LZ}\rightarrow \mathcal{LZ}^{\prime }+\mathcal{S}^{\prime }$ and $\mathcal{LZ}^{\prime }\rightarrow \mathcal{LZ}^{\prime \prime }+\mathcal{S}^{\prime \prime }$ in the form of spectral peaks with frequencies corresponding to  $\mathcal{LZ}$,  $\mathcal{LZ}^\prime$, $\mathcal{LZ}^{\prime \prime }$, $\mathcal{S}^{\prime}$ and $\mathcal{S}^{\prime \prime }$ waves. One measures that $\omega _{\mathcal{LZ}}^{D}\simeq0.979\omega _{p},$ $\omega _{\mathcal{LZ}^{\prime }}^{D}\simeq1.039\omega _{p},$ $\omega _{\mathcal{LZ}^{\prime \prime }}^{D}\simeq0.987\omega _{p},$ $\omega _{\mathcal{S}^{\prime }}^{D}\simeq0.059\omega _{p}$, and $\omega _{\mathcal{S}^{\prime \prime }}^{D}\simeq0.048\omega _{p}$. The three-wave resonance conditions $\omega _{\mathcal{LZ}}^{D}=\omega _{\mathcal{LZ}^{\prime }}^{D}-\omega _{\mathcal{S}^{\prime}}^{D} $ and $\omega _{\mathcal{LZ}^{\prime }}^{D}=\omega _{\mathcal{LZ}%
^{\prime \prime }}^{D}+\omega _{\mathcal{S}^{\prime \prime }}^{D}$ are satisfied with good accuracy. Furthermore, the squared cross-bicoherence $b_c^2$ calculated within the same time interval for the triplet $(E_{\parallel },\delta {n}_{i},E_{\parallel })$ provides that $b_{c}\simeq0.6$ and $b_c\simeq0.63$ at the  frequencies corresponding to the first and second cascades, respectively (see Figure \ref{fig8}h and its caption). 

Figures \ref{fig8}a-h confirm the occurrence of decay in the recorded waveforms. Moreover, the late increase of  $E_\perp(t)$ in  Figure \ref{fig8}a  results from ESD transporting wave energy to smaller, more oblique wavevectors  (\cite{Polanco2025b}). At  $\omega_pt\gtrsim 8500$, $\mathcal Z$-mode waves appear during the last stage of ESD (\cite{Polanco2025a}), when ion acoustic waves' frequencies satisfy $\omega^D\lesssim0.035\omega_p$. However, clear cross-bicoherence can be hardly evidenced at this time, due to the rarity of phase-coherent wavepackets in the developed wave turbulence. Jointly, $\mathcal O$-mode waves are generated at $\omega_pt\gtrsim5000$ via EMD (\cite{Krafft2024}), with low-frequency signatures at $\omega^D_{\mathcal S}\simeq0.025\omega_p$. Note that the generation of electromagnetic waves via the EMD process in a weakly magnetized plasma will be thoroughly treated in a forthcoming paper.

\begin{figure*}[!htb]
    \centering
    \includegraphics[width=0.8\linewidth]{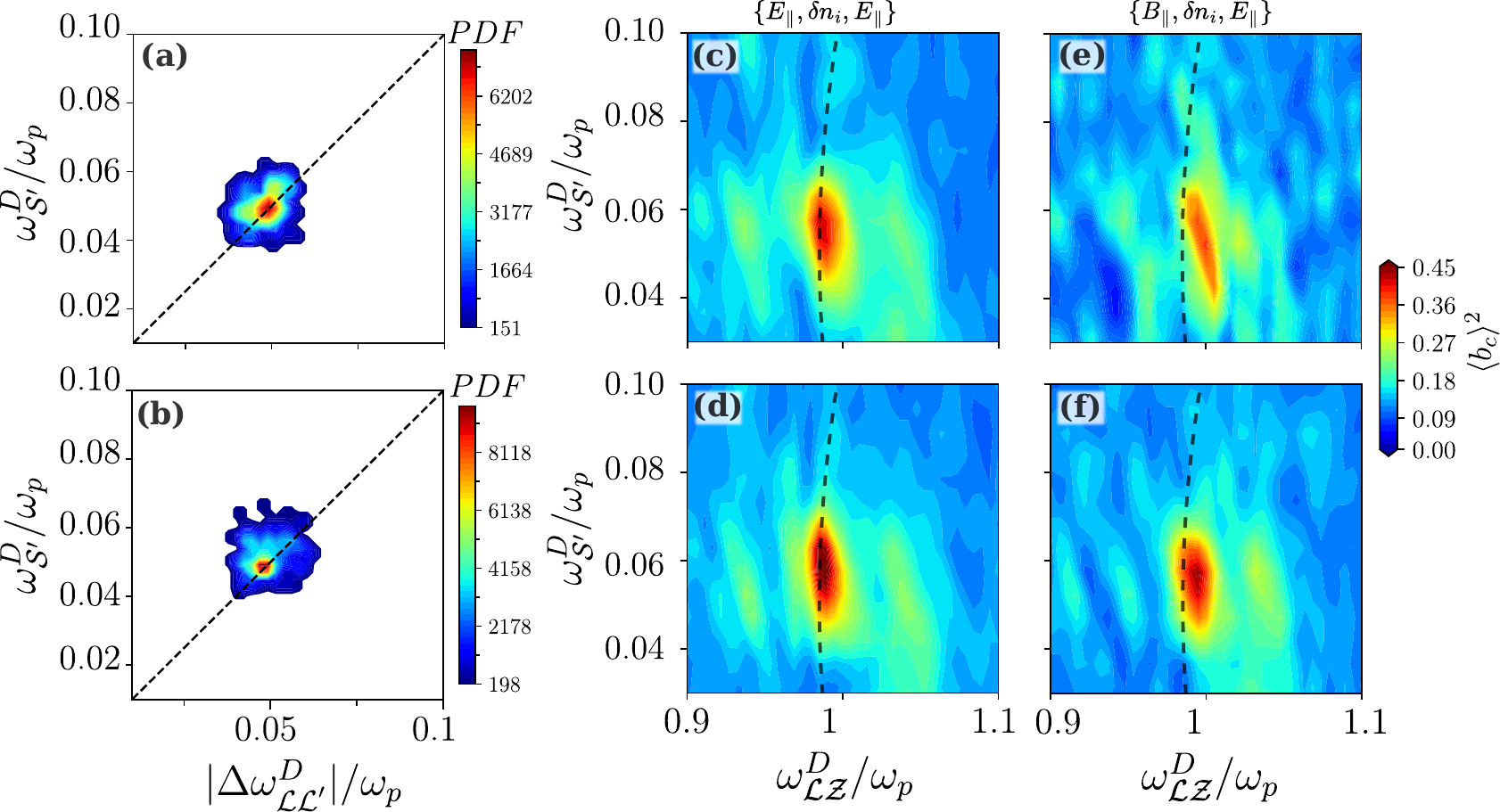}
    \caption{(a-b) Wave distributions in the plane ($\left\vert\Delta\omega_{\mathcal{LL}^{\prime}}^{D}\right\vert /\omega_p$, $\omega_{\mathcal{S}^{\prime}}^{D}/\omega_p$), for  $N_s$ selected spectra out of a set of $256 $; (a) : $N_s=154$, $\omega_{c}/\omega_{p}=0.005$;   (b) : $N_s=150$,  $\omega_{c}/\omega_{p}=0.07$; both distributions are calculated within the time interval $\Delta T=[1000,6000]\omega_p^{-1} $. The dashed lines represent the ESD resonance condition $\omega_{\mathcal{LZ}^{\prime}}^{D}-\omega_{\mathcal{LZ}}^{D}=\omega_{\mathcal{S}^{\prime}}^{D}$. (c-d) Square cross-bicoherence $\langle b_c\rangle^2$ averaged on $N_{bc}$ selected waveforms consistent with ESD occurrence within $\Delta T$,  in the map ($\omega_{\mathcal{LZ}}^{D}/\omega_p,\omega_{\mathcal{S}^{\prime}}^{D}/\omega_p$), for the triplet ($E_{\parallel},\delta n _{i},E_{\parallel}$)  ($N_{bc}=60$ (c) and $N_{bc}=64$ (d)). (e-f) Square  cross-bicoherence $\langle b_c\rangle^2$  averaged on $N_{bc}$  selected waveforms  within $\Delta T$,  in the map ($\omega_{\mathcal{LZ}}^{D}/\omega_p,\omega_{\mathcal{S}^{\prime}}^{D}/\omega_p$), for the triplet ($E_{\parallel},\delta n_{i},B_{\parallel}$)  ($N_{bc}=9$ (e) and $N_{bc}=35$ (f)). (c-f) : the  dashed lines represent the theoretical curves described in Figure \ref{fig3}. (Upper row) :  $\omega_{c}/\omega_{p}=0.005$.  (Bottom row) :  $\omega_{c}/\omega_{p}=0.07$.}
    \label{fig9}
\end{figure*}

\subsection{Magnetic signatures of decaying $\mathcal{LZ}$ waves }

Figures \ref{fig9}a,b show wave distributions in the map   ($|\Delta \omega_{\mathcal{LL}^{\prime }}^{D}|/\omega_p, \omega _{\mathcal{S}^{\prime }}^{D}/\omega_p)$,  for  $N_s=154$ (a) and $N_s=150$ (b) selected spectra  featuring a multi-peak structure (out of a set of $256$), in  magnetized plasmas with  $\omega _{c}/\omega _{p}=0.005$ and $0.07$, respectively. The corresponding  squared average cross-bicoherence $\langle b_c\rangle^2$ is presented in Figures \ref{fig9}c-f, for the triplets $(E_\parallel,\delta n_i,E_\parallel)$ (c,d) and $(B_\parallel,\delta n_i,E_\parallel)$ (e,f), respectively, within the time domain $1000\lesssim \omega _{p}t\lesssim 6000$.  In both cases, bicoherence maxima up to  $\langle b_c\rangle\simeq0.7$ are aligned along  the  theoretical curves representing the three-wave resonance conditions (compare Figures \ref{fig3}c and \ref{fig9}c,d).  The 
number of waveforms (out of a set of 256)
exhibiting clear ESD signatures closely matches that of the unmagnetized plasma case ($N_{bc}=50$ for $\omega_c=0$ and $N_{bc}=60$  for $\omega_c\neq0$). Furthermore, ESD occurs for $\omega_c=0$ and  $\omega_c>0$  at similar frequencies and wavevectors, as  similar zones on the  ($\omega_{E_{\parallel}}^{D},\omega_{\delta n_i}^{D}$) and  ($\omega_{B_{\parallel}}^{D},\omega_{\delta n_i}^{D}$) planes present high values of $\langle b_c\rangle^2$. This shows that the efficiency of this process at large $k$ is only slightly affected by the weak plasma magnetization. This was already studied in a previous work using a global approach (\cite{Polanco2025a}); however, the methodology presented here can be applied to analyze  actual $\mathcal{LZ}$ waveforms observed by satellites in the solar wind.

As $\mathcal{LZ}$ waves are weakly (strongly) magnetized at  large (small) $k$-scales (see also Figures \ref{fig9}e,f), they generate through ESD waves with substantial magnetic components. This conclusion is also valid for higher order cascades, which are however not relevant at early times $\omega_pt\lesssim 6000$. This makes the number of waveforms consistent with ESD to be smaller for the triplet $(B_\parallel,\delta n_i,E_\parallel)$ than for  $(E_\parallel,\delta n_i,E_\parallel)$. As plasma magnetization decreases, the spectral region of waves with large magnetic signatures --- around $k_*\lambda_D=(v_T/c)/({1+\omega_p/\omega_c})^{-1/2}$--- also decreases, and  larger times are needed for decay to reach such small-$k$ region.
 This explains why fewer waveforms consistent with ESD are observed  for the triplet $(B_\parallel,\delta n_i,E_\parallel)$ at  $\omega_c/\omega_p=0.005$ ($N_{bc}=9$ in Figure  \ref{fig9}e) compared to $\omega_c/\omega_p=0.07$ ($N_{bc}=35$ in Figure  \ref{fig9}f). 

\begin{figure*}[!htb]
    \centering
    \includegraphics[width=0.85\linewidth]{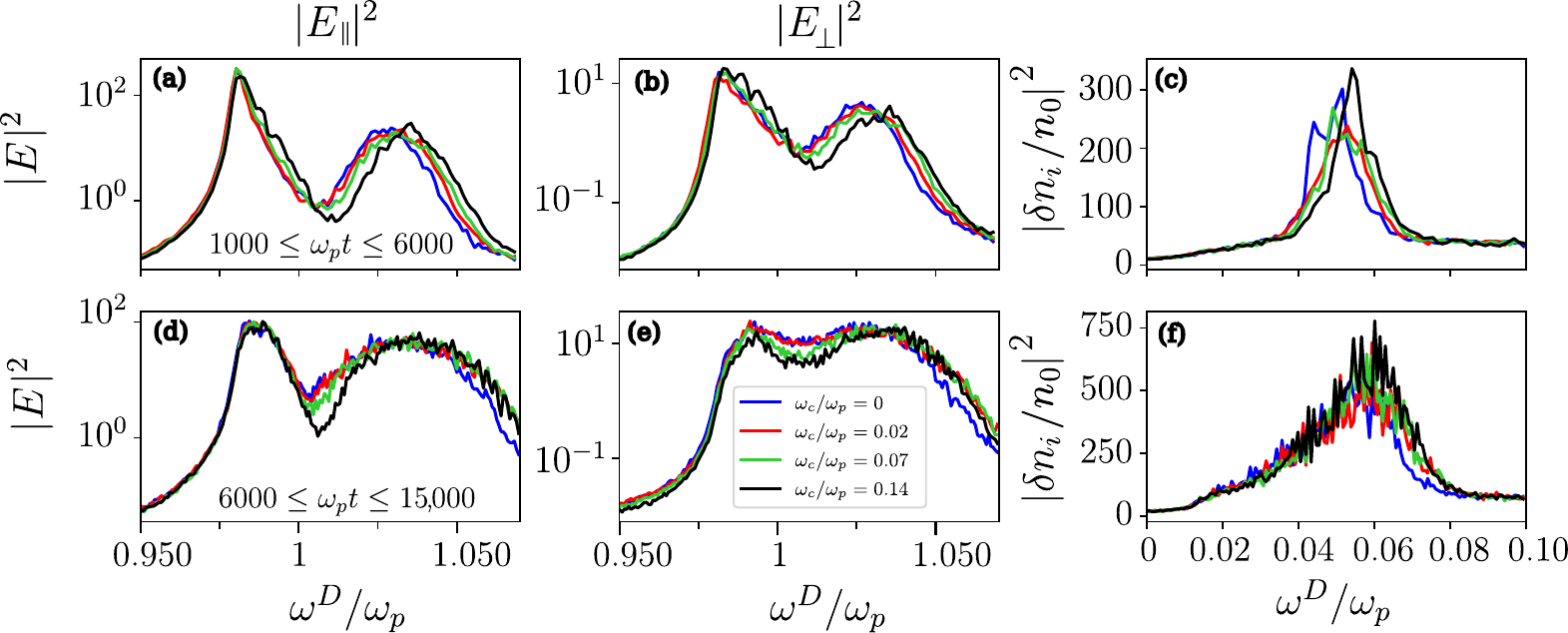}
    \caption{Energy spectra averaged over $N_s=256$ waveforms, as a function of  $\omega^{D}/\omega_{p}$, for different magnetization ratios $\omega_{c}/\omega_{p}=0, 0.02, 0.07,$ and $ 0.14$ (see the legend), in the time ranges  $1000\lesssim\omega_{p}t\lesssim6000$ (top row)  and $6000\lesssim\omega_{p}t\lesssim15,000$ (bottom row). (a,d) Parallel $\left\langle |E_{\parallel}|^{2}\right\rangle $ and (b,e) perpendicular $\left\langle |E_{\perp}|^{2}\right\rangle $ electric energy spectra. (c,f) Ion acoustic wave spectrum $\left\langle |\delta n_{i}/n_{0}|^{2}\right\rangle $. (a-b, d-e): logarithmic scales; (c,f) : linear scales. All variables are in arbitrary units. }
    \label{fig10}
\end{figure*}

\subsection{Wave turbulence spectra versus magnetization}

Figures \ref{fig10}a-f show the spectral distributions of $\left\langle |E_{\parallel
}|^{2}\right\rangle $, $\left\langle |E_{\perp }|^{2}\right\rangle $,  and $\langle |\delta n_i/n_0|^2\rangle$,  averaged over $256$ waveforms,  in the time ranges $1000\lesssim \omega _{p}t\lesssim
6000$ (a-c) and $6000\lesssim \omega _{p}t\lesssim
15,000$ (e-f), for different magnetization ratios. When $\omega_{c}/\omega _{p}$ increases, both electric energies decrease within the frequency range $0.99\omega _{p}\lesssim \omega
^{D}\lesssim 1.02\omega _{p}$ (Figures \ref{fig10}a,d), confirming the impact of magnetization on  $\mathcal{LZ}$  waves of small $k$  (i.e. at $\omega^D\sim\omega_p$). Indeed, $\mathcal{LZ}$  wave energy cannot penetrate the region surrounding $k\sim0$ (\cite{Polanco2025a}).
In this frequency range only, and mostly at advanced times,  $\left\langle |E_{\perp }|^{2}\right\rangle $  significantly exceeds   $\left\langle |E_{\parallel
}|^{2}\right\rangle $ when  $\omega_{c}/\omega _{p}\leq0.14$, so that perpendicular energy $ |E_{\perp }|^{2}$ increases at small $k$-scales, when  $\mathcal{LZ}$ waves become quasi-electromagnetic (see also \cite{Polanco2025b}).
On the other hand,  the electric energy spectra  broaden significantly over time (compare Figures \ref{fig10}a-b and  \ref{fig10}d-e), due to the beam  exciting $\mathcal{LZ}$ waves of larger $k$ during its relaxation and the redistribution of energy over smaller $k$-scales by ESD cascades. In particular, a small shift toward larger frequencies can be observed at $\omega^D\sim1.03\omega_p$ for the largest $\omega_c/\omega_p\simeq0.14$, which is also visible at $\omega^D\sim0.055\omega_p$ for the ion acoustic waves produced by ESD (Figures \ref{fig10}c,f). Note that \cite{Krauss-Varban1989} calculated the linear growth rate of $\mathcal{LZ}$ waves excited by a beam and observed also  a slight $k$-shift with growing $\omega_c/\omega_p$.  For  $\omega_c/\omega_p<0.14$,   features of low-frequency spectra are comparable  to those of  Figure \ref{fig2}, showing the weak impact of magnetization. The time evolution of spectra at any $\omega_c/\omega_p$ shows  the broadening of $\langle |\delta n_i/n_0|^2\rangle$ and its extension toward  larger  (smaller) frequencies over time, due to beam relaxation (higher order decay cascades).  

\section{Discussion and conclusion}
Our study investigates wave processes occurring during type III solar radio bursts by using large-scale and long-term PIC simulations. By reproducing waveforms that closely match spacecraft observations in the solar wind, we identify main nonlinear and linear wave phenomena. Notably, we observe  wave-wave interaction processes such as electrostatic and electromagnetic decay (ESD and EMD), alongside linear transformations of turbulent electrostatic wavepackets on random density fluctuations —including reflection, refraction, trapping, tunneling and, crucially, the linear mode conversion at constant frequency (LMC). A key finding is the dynamic interplay and competition between these processes. Specifically, we demonstrate that wave scattering and subsequent  LMC ---which strongly shape the $\mathcal{LZ}$ wave turbulence evolution--- can initiate nonlinear processes much earlier than expected in plasmas without random density fluctuations.
 
Electrostatic decay (ESD) is widely regarded as the most efficient nonlinear wave-wave interaction process in the solar wind and a cornerstone in understanding type III solar radio bursts. While electromagnetic decay (EMD) and linear mode conversion (LMC) have yet to be unambiguously observed in the solar wind, simultaneous detections of Langmuir and acoustic waves consistent with the ESD mechanism have been reported in previous studies (\cite{Henri2009},  \cite{Malaspina2011}, \cite{GrahamCairns2013b}, \cite{Kellogg2013}). Analyses of frequency spectra have been conducted to identify three-wave resonance conditions, leveraging in some cases extensive waveform datasets and statistical methods. However, these studies have not provided definitive evidence for ESD’s occurrence, due to very low statistics and/or to lack of phase coherence diagnostics between waves.  
We advance this approach by statistically analyzing a large dataset of PIC-simulated waveforms under a controlled framework. In addition to frequency spectra and resonance conditions, our study  integrates bicoherence diagnostics and analysis of magnetic energy fluctuations, ion density waveforms under varying levels of density turbulence, and three-dimensional electric fields. This framework not only aligns PIC simulation results with solar wind observations but, being more comprehensive than space-recorded waveforms, also demonstrates  the stimulation of nonlinear processes by LMC and enables the evidence of ESD occurrence under different plasma conditions.
In homogeneous plasmas (randomly inhomogeneous plasmas with $\Delta N\gtrsim 3(v_T/v_b)^2$), our analysis reveals three-wave ESD resonance conditions in 60\% (20\%) of simulated waveforms, with high bicoherence levels observed  only in 20\% (7\%).  

Despite some discrepancies, our findings align with space-based waveform analyses.
\cite{GrahamCairns2013b} reported that about 40\% of the analyzed waveforms exhibit spectral peaks ---consistent with ESD resonance conditions---, a value intermediate between our results for homogeneous (60\%) and ransomly inhomogeneous (20\%) plasmas. This discrepancy can be explained in light of previous results.
Both \cite{Malaspina2011} and \cite{GrahamCairns2013b} linked observed  high polarization ratios $F=|E_\perp|^2/|E|^2$  to small-$k$ waves whose origin was debated. They also  observed that  polarization ratios  increase with $v_b/c$, presenting a  sharp rise at  $v_b/c\simeq0.08$. \cite{Polanco2025b} explained this growth by the presence of random density fluctuations that strongly interact with $\mathcal{LZ}$ waves when  $\Delta N\gtrsim 3(v_T/v_b)^2$. The 40\% of waveforms consistent with ESD found by \cite{GrahamCairns2013b} were produced by electron beams  with velocities above and below $v_b/c\simeq0.08$, i.e. in plasmas where  $\Delta N\lesssim 3(v_T/v_b)^2$ and $\Delta N\gtrsim 3(v_T/v_b)^2$, respectively. This explains the global occurrence of 40\% that lies between the values of 20\% and 60\% provided by our simulations.  

\cite{GrahamCairns2013b} and \cite{Kellogg2013} found events with two ESD cascades but noted that spectra featuring more than two cascades are extremely rare,  occurring only in the presence of unusually intense $\mathcal{LZ}$ waves. Our results support this conclusion, and we demonstrate that, while uncommon, such ESD cascades can indeed occur.

This paper establishes a direct correspondence between  PIC simulation and spacecraft-observed waveforms during type III solar radio bursts. By analyzing waveforms recorded under solar wind conditions, this work offers methodological guidance for reproducing and extending these results. The findings facilitate direct comparison between spacecraft observations and simulation benchmarks, enabling the identification of observational gaps and the improvement of detection capabilities. 

\section{Acknowledgements}
This work was granted access to the HPC computing and storage resources under the allocation 2023-A0130510106 and 2024-A017051010 made by GENCI. This research was also financed in part by the French National Research Agency (ANR) under the project ANR-23-CE30-0049-01.  C.K. thanks the International Space Science Institute (ISSI) in Bern through ISSI International Team project No. 557, Beam-Plasma Interaction in the Solar Wind and the Generation of Type III Radio Bursts. C. K. thanks the Institut Universitaire de France (IUF). F. J. P. R. thanks the \'Ecole Universitaire de Recherche (EUR) Plasma Science.

\printbibliography

\end{document}